\documentclass[12pt, draftclsnofoot, onecolumn]{IEEEtran}
\normalsize
\usepackage{amsmath}
\usepackage{amsmath,amsthm,amssymb}
\usepackage{color,soul}
\usepackage{amsfonts}
\usepackage{epsfig}
\usepackage{amssymb}
\usepackage{cite}
\usepackage{subfigure}
\usepackage{multirow}
\usepackage{rotating}
\usepackage{graphicx}
\usepackage{tabularx}
\usepackage{array}
\usepackage{amsfonts}
\usepackage{graphicx}  
\usepackage{graphics}  
\usepackage{epsfig}
\usepackage[all]{xy}
\usepackage{tabularx}
\usepackage{amsmath}
\usepackage{amsfonts}
\usepackage{epstopdf}
\usepackage{amssymb}
\usepackage{cite}
\usepackage{subfigure}
\usepackage{multirow}
\usepackage{rotating}
\usepackage{graphicx}
\usepackage{tabularx}
\usepackage{array}
\usepackage{setspace}
\usepackage{morefloats}
\usepackage{xcolor}
\usepackage{float}
\usepackage[Euler]{upgreek}
\usepackage{mathtools}
\usepackage{algorithm} 
\usepackage[algo2e]{algorithm2e}

\usepackage[mathscr]{eucal}
\ifCLASSINFOpdf
\else
\fi

\begin{document}
%
\title{AI-Based and Mobility-Aware Energy Efficient Resource Allocation and Trajectory Design for NFV Enabled Aerial Networks}
%
%
%

\author{Mohsen~Pourghasemian,~Mohammad~Reza~Abedi,~Shima~Salarhosseini,~Nader~Mokari,~\IEEEmembership{Senior~Member,~IEEE,} Mohammad~Reza~Javan,~\IEEEmembership{Senior~Member,~IEEE~}and~Eduard~ A~Jorswieck,~\IEEEmembership{Fellow,~IEEE}

\thanks{ M. Pourghasemian, Mohammad. R. Abedi, S. Salarhosseini and N. Mokari are with the Department of Electrical and Computer Engineering, Tarbiat Modares University, Tehran, Iran e-mail: (\{Mohammadreza$\_$abedi, shima.salarhosseini and nader.mokari\}@modares.ac.ir), Mohammad R. Javan is with the Department of Electrical and Robotics Engineering, Shahrood University of Technology, Shahrood, Iran e-mail: (javan@shahroodut.ac.ir) and Eduard A. Jorswieck is with  TU Braunschweig, Department of Information Theory and Communication Systems, Braunschweig, Germany (jorswieck@ifn.ing.tu-bs.de). This work was supported by the joint Iran national science foundation (INSF) and German research foundation (DFG) under grant No. 96007867.}}


%
%

\markboth{}%
{Submitted paper}
%



\maketitle

\begin{abstract}
In this paper, we propose a novel joint intelligent trajectory design and resource allocation algorithm based on user's mobility and their requested services for unmanned aerial vehicles (UAVs) assisted networks, where UAVs act as nodes of a network function virtualization (NFV) enabled network. Our objective is to maximize energy efficiency and minimize the average delay on all services by allocating the limited radio and NFV resources. In addition, due to the traffic conditions and mobility of users, we let some Virtual Network Functions (VNFs) to migrate from their current locations to other locations to satisfy the Quality of Service requirements. We formulate our problem to find near-optimal locations of UAVs, transmit power, subcarrier assignment, placement, and scheduling the requested service's functions over the UAVs and perform suitable VNF migration. Then we propose a novel Hierarchical Hybrid Continuous and Discrete Action (HHCDA) deep reinforcement learning method to solve our problem. Finally, the convergence and computational complexity of the proposed algorithm and its performance analyzed for different parameters. Simulation results show that our proposed HHCDA method decreases the request reject rate and average delay by 31.5\% and 20\% and increases the energy efficiency by 40\% compared to DDPG method.\\
\end{abstract}


%
\IEEEpeerreviewmaketitle

\section{Introduction}
%
%
%
%
\subsection{Motivation}
{Nowadays}, daily life seriously depends on data networks and costumers which expect to have services in any circumstances. For instance, gathering of large number of people at a short period of time in one place, sport matches, or heavy traffics in the crowded crossover, causes an overload to the local network. Moreover, during the natural disaster in which a lot of infrastructures are destroyed and the conventional Base Stations (BSs) do not work properly, users can hardly access to the network. One of the most effective approaches to overcome the aforementioned challenges is to deploy the Unmanned Aerial Vehicle assisted (UAV-assisted) networks, in which UAVs can play the roles of BSs, in addition to their abilities to move and create an ad-hoc network \cite{s1}. However, the available resources in UAVs are limited and they should be properly managed so that they can provide better performance. In an UAV-assisted network, the locations of UAVs and the way they change their locations, affect the power consumption and the Quality of Service (QoS) \cite{s2}.

Network function virtualization (NFV) is one of the key technologies for the next generation cellular networks, which can reduce the deployment cost of network functions \cite{N1}. In a traditional way, network functions (NFs) are implemented on the dedicated hardwares. This approach not only lacks of flexibility, but also it is expensive \cite{N2}. Instead, NFs can be implemented as software on general purpose hardware and do the same functionality such as firewall or traffic monitor. This can help operators to quickly provide the new NFs for mobile networks. A network service consists of a service chain of two or more Virtual Network Functions (VNFs) which should be executed on different nodes. In other words, the VNFs are deployed over some nodes that can be linked to create a service chain in response to a demand \cite{NU1,NU2}. Changing the locations of mobile users and their service requests will oblige NFV enabled networks to perform VNF migration in order to satisfy the network QoS requirements \cite{N2}. In other words, considering a real environment, users are mobile and their requested services are dynamic hence, to guarantee QoS and reduce latency, we need to do proper migration planning where the users receive the requested services seamlessly. In addition, new requirements of emerging services like virtual reality have high data rate\cite{virtualreality}. To overcome this, we can use a new spectral efficient non orthogonal multiple access (NOMA) technology like power domain NOMA (PD-NOMA) which can improve the data rate of users. Compared with previous orthogonal multiple access (OMA) technologies like orthogonal frequency-division multiple access (OFDMA), PD-NOMA can support more users and offer higher data rates\cite{NOMA-SPECFF}.

 
 The optimal resource allocation, function placement, and scheduling in NFV-enabled networks need more attention since the NFV technology provides more flexibility to the resource limited wireless networks. On the one hand, most of the resource allocation and VNF placement problems are non-convex and NP-hard, therefore, it is very difficult to obtain the optimal solutions by traditional methods, especially in large scale networks. In addition, as the number of optimization variables increase, the traditional methods may not achieve the sub-optimal solution and their performance complexity increases exponentially. Also, the traditional methods must be solved every time the network parameters change, therefore it is impractical to use the traditional methods for such complex problems. To tackle these challenges, Artificial Intelligence (AI)-based algorithms can be effectively used to solve such optimization problems. One of the best AI-based solutions is Reinforcement Learning (RL) in which an agent observes the environment, takes an action, and based on the received reward chooses a strategy to reach the efficient policy \cite{rl}. However, RL cannot operate well in the large environment because it needs a huge amount of resources such as memory and processor. In this regard, the Deep RL (DRL) methods are introduced which utilize deep neural networks to process the learning procedure \cite{drl}. DRL methods can provide solutions for modern networks which arise a large environment with high computational complexity. In addition, since the centralized single-agent methods lead to large signaling overhead in large networks, DRL based solutions can be used in a decentralized multi-agent manner which become more popular in recent networks such as UAV-assisted networks\cite{drlm}. In this approach, each agent takes its own decision independently of the others. Therefore, it is not required to obtain global information and can reduce the signaling overhead in large networks \cite{maddpg}.

\subsection{Literature Review}

\subsubsection{UAV Deployment and Radio Resource Allocation}
The authors in \cite{3_arx} considered an UAV-assisted network in which the UAVs operate as on-board edge servers and collect data from Internet of Things (IoT) nodes. They proposed a two-dimensional path planing and an optimal transmit power for UAVs by utilizing Q-learning in which the UAVs were trained by a centralized cloud with the amount of collected data as the objective function. A movement and deployment design for multiple UAVs were considered in \cite{7} where the authors considered the effect of user mobility in their design, and proposed an AI-based three-dimensional path planning by utilizing Q-learning. The aim of their design was to maximize the user's quality of experience as a function of transmission rate. A trajectory design for UAVs was developed in \cite{8} by considering both ground and aerial users. The aim in \cite{8} was to maximize the number of serving users by proposing a double Q-learning based algorithm.
The authors in \cite{4} considered a vehicular control network and utilized an UAV to improve their network performance. They utilized a deep learning algorithm based on Deep Deterministic Policy Gradient (DDPG) for resource allocation and UAV trajectory design. A resource allocation design was proposed in \cite{13}, in which the data rate of each UAV was maximized and Markov Decision Process (MDP) was adopted to their problem in order to solve it using Q-learning method. In \cite{19}, the authors minimized the transmit power and power consumption per movement by an optimal placement of UAVs. They utilized a novel machine learning algorithm based on Gaussian mixture model and a weighted expectation maximization algorithm to deploy UAVs in a not congested way with the minimum power consumption and movement cost. 
\subsubsection{NFV Management}
The authors in \cite{da} investigated the NFV enabled network, and founded the optimal function scheduling to minimize the processing and transmission delay. They solved the problem based on Genetic algorithm. In \cite{zakeri}, the authors proposed an algorithm for scheduling and VNF placement to minimize the number of virtual machines and the total transmit power. The authors adopted NFV in \cite{E2E}, in which they provided Tactile Internet (TI) service. They considered joint VNF placement and radio resource allocation, while guaranteeing the required delay for TI service by finding the optimal transmit power and function scheduling. In \cite{fp6}, the authors considered a function placement problem in a cloud, where their aim was to minimize the latency by considering the capacity limitation, node availability, and tolerated delay of the services. The authors developed an AI-based solution for function placement in order to minimize the number of utilized nodes and links. They exploited deep Q-learning as a solution to their problem \cite{dqn}. According to the literature, non of the previous works considered the function placement on UAVs to serve the mobile users. 

\subsubsection{VNF Migration} 
The authors in \cite{Hirayama} considered VNF migration scheduling problem based on an encoder-decoder recurrent neural network.
In \cite{q2}, the authors considered impact of user's mobility on VNF placement in a cellular network, aiming to minimize the cost of VNF migration.
In \cite{q3}, the authors investigated the mobility of users that may cause the NFV enabled network to re-route the VNF placement to handle the requested service.	
In \cite{99333}, the authors combined mechanism for prediction and migration together to optimize the cost of VNFs migration and guarantee the quality of services that may be decreased by resource limitations. They proposed a real-time VNF migration algorithm based on deep belief network to predict future resources requirements. They investigated bandwidth utilization and migration problems jointly.
\textcolor{black}{\subsubsection{UAV Trajectory Design }
The authors in\cite{gholami2020joint} proposed a framework for joint UAV placement and route optimization
in a multi-hop UAV relaying communications system, taking into account the mobility
of the ground nodes, the UAVs mobility constraints, and the UAVs propulsion energy consumption. In \cite{diao2019joint} the trajectory, task data, and computing resource allocation had been joinly optimized to minimize the energy consumption among UAVs. The resource allocation problem for multiple users were considered in the UAV MEC-enabled network in \cite{hu2018joint}, in which computation offloading, UAV trajectory and user scheduling were jointly studied. Authors in \cite{hu2020cooperative} proposed an UAV-assisted and distributed sense-and-send protocol in which a decentralized DRL based trajectory design were developed to ensure coverage and user fairness to minimize the age-of-information (AoI) when the UAVs execute sensing tasks through cooperative sensing and transmission.}
\subsubsection{Deep Reinforcement Learning based solution}
AI-based methods are promising solutions for solving the problems which are NP-hard and non-convex \cite{ql0}. However, depending on the problem, the state-of-the-art methods such as Deep Q Network (DQN) and DDPG may lose to perform optimally due to their restrictions. For instance, DQN cannot perform optimally in the environment with continuous actions\cite{ql0}, also DDPG cannot be directly used for the environment that contains discrete actions \cite{quantizeDisc}. On the other hand, there are environments which have both discrete and continuous actions simultaneously, and for which the parameterized actions space based DRL \cite{hybridDC} methods are perfectly suited. However, in some situations, like as our system model, the discrete and continuous actions in the environment are dependent to each other. For instance, the discrete actions should be taken based on the continuous ones and the existing DRL methods cannot be directly used to solve such a hybrid environment.

From the above literature review, there is not an UAV-assisted network which considers UAVs as NFV enabled nodes so that the VNF placement and migration approach with the limited radio and NFV resources are performed by the UAVs, aiming to satisfy the QoS of the mobile users. In the proposed VNF migration enabled UAV-assisted network, the locations of UAVs not only should be changed related to the locations of users, but also should be determined based on the requested services of users. Moreover, the limitation of power budget and available bandwidths between UAVs may results to increase the delay of services. Another challenge for this kind of networks is the Central Processing Unit (CPU) limitation for UAV nodes, which causes that they can run only a few number of functions. 
In some environments, like UAV-assisted and migration enabled NFV networks, as VNF placement and radio resource allocation have dependent continuous and discrete values, DRL methods like DDPG and DQN cannot optimally perform due to their limitation for the action selection. Therefore in this paper, we propose a new Hierarchical Hybrid Continuous and Discrete Action (HHCDA) DRL method which considers joint discrete and continuous actions to find near-optimal solution for our proposed problem.  

\subsection{Our Contributions}
In this paper, we consider an  UAV-assisted network, where the UAVs are utilized as NFV nodes to play BS roles. 
We formulate our problem as an optimization problem, in which we aim to jointly maximize the Energy Efficiency (EE) and minimize services' delay by finding optimal transmit power, subcarrier assignment, UAVs' location, service function chaining, function placement and scheduling subject to network and resource constraints. According to the discussion above, the primary contributions of our paper are listed as follows:
\begin{itemize}
	\item{
		To the best of our knowledge, the existing works in the literature did not utilize migration enabled VNF placement and scheduling over UAVs. We assume that UAVs are the NFV enabled nodes on which the VNFs can be placed in an optimal way. Also, we consider that VNF migration can take place due to user's mobility and their diverse service requests.}
	\item{
		We propose a novel trajectory design algorithm for UAVs based on joint user's mobility and service requests, in which the EE of 
		UAVs are maximized and the delay of the requested services is minimized.
	}
	\item{
		With the help of two efficient DRL methods as DQN and DDPG, 
		HHCDA DRL method is proposed to solve our problem which can handle the continuous and discrete actions, including transmit power allocation, subcarrier assignment, UAV trajectory design, VNF placement and scheduling.
	}
	
	\item{By conducting simulation, we compare our proposed HHCDA solution with the baseline DRL methods from performance computational complexity and convergence perspectives. Simulations validate that our proposed DRL method outperforms \textcolor{black}{the benchmark approaches from the request reject rate (RRR), average delay, and energy efficiency point of view by 35\%, 19\%, and 26\%, respectively.} .}
\end{itemize}
\subsection{Organization and Notation}
The remainder of this paper is organized as follows: In Section II, the system model is presented. In Section III, the optimization problem is formulated, and the elements of learning based algorithm are determined in Section IV. Simulation results and their analysis are discussed in Section V, and finally in Section VI, the conclusion is presented.

Notations: 
Vectors and matrices are denoted by boldface small and big letters, respectively. $\nabla_aQ$ determined the gradient of $Q$ over $a$, and Pr$(J=J')$ is used to show the probability when $J=J'$. $\mathbb{E}_s\{.\}$ is the expectation function over $s$, and $|.|$ denotes the number of elements in the set.
\section{System Model}
We consider an UAV-assisted network, where each UAV serves a number of users. There is a set of UAVs as $\mathcal{U}$ indexed by $u$, where $|\mathcal{U}|=U$ denotes the number of UAVs. Each UAV is equipped
with a single antenna. A time set $\mathcal{T}$ is indexed by $t$, where $|\mathcal{T}|=T$ denotes the number of time slots. The time slots have equal duration of $\iota$. The location of UAV $u$ at time slot $t$ is denoted by $\check{q
	}_u(t)=\big(\check{x}_{u}(t), \check{y}_{u}(t), \check{z}_{u}(t)\big)$ where $\check{x}_{u}(t)\in\mathbb{R}, \check{y}_{u}(t)\in\mathbb{R}$, and $\check{z}_{u}(t)\in\mathbb{R}$ are horizontal, vertical, and altitude coordinates of UAV $u$ at time slot $t$, respectively. We assume that the maximum UAV's velocity is $W^{\text{max}}$. Therefore, each UAV can travel at maximum distance as $D^{\text{max}}=W^{\text{max}}\iota$ between two adjacent slots. To avoid the collision, the minimum safe distance, $D^{\text{min}}$ must be kept between UAVs. Thus, for UAV $u$, the following mobility constraint must be satisfied to restrict the its trajectory \cite{ji2020joint}:
\begin{equation}
\text{C1:} \left\| \check{q}_u(t+1)-\check{q}_u(t) \right\| \leq D^{\text{max}}, \text{C2:} \left\| \check{q}_u(t)-\check{q}_{u'}(t) \right\| \geq D^{\text{min}}, \forall u\neq u',
\end{equation}
where $\left\|.\right\|$ denotes the Euclidean norm. Constraint C1 denotes that the maximum
movement of the UAVs at each time slot is restricted by its maximum flying speed. Constraint C2 indicates that the minimum distance between two UAVs should not be less than the minimum value, $D^{\text{min}}$. The assigned user set to UAV $u$ is denoted by $\mathcal{K}_u$ with $|\mathcal{K}_u|={K}_u$. Each user can be served by only one UAV at a time and is equipped
with a single antenna. The set of assigned users to all UAVs is denoted by $\mathcal{K}_{\text{Tot}}= \bigcup_{u=1}^{U}\mathcal{K}_u$ with $K_{\text{Tot}}=|\mathcal{K}_{\text{Tot}}|$. We construct a set of total nodes which contains the users and the UAVs as $\mathcal{N} = \{\mathcal{U},\mathcal{K}_{\text{Tot}}\}$ with $U+K_{\text{Tot}} = N$. The location of user $k$ at time slot $t$ is represented by $\tilde{q}_{k}(t)=(\tilde{x}_{k}(t),\tilde{y}_{k}(t))\in\mathbb{R}$ which are the horizontal and vertical coordinates of user $k$, respectively. Since the users change their locations at each time slot, we consider the random walk model for their mobility \cite{rw}. The users move uniformly in any direction with a random speed between 0 and $V^{\text{max}}$. 

\begin{figure}
	\center
	\includegraphics[scale = .8]{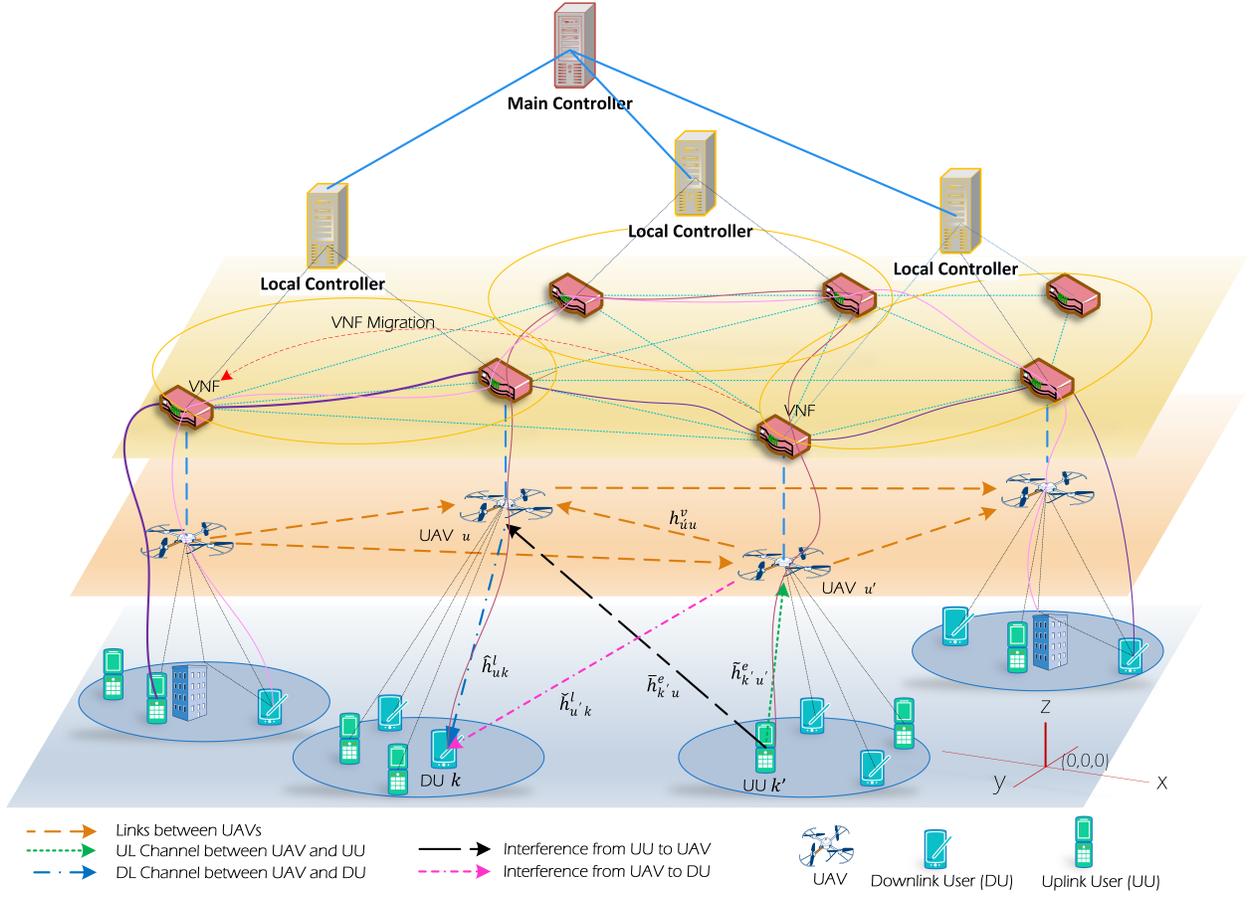}
	\center
	\caption{The considered system model.}\label{u2u}
\end{figure}

We assume that the network uses three separated bands for links between UAV-UAV, UAV-User, and User-UAV so that there is no overlap between these links. Each bandwidth is divided into several orthogonal subcarriers. We assume that the bandwidth of UAV-UAV communication, $\hat{B}$ is divided into a set of $\mathcal{V}$ subcarriers each with bandwidth $\hat{B}_1$ denoted by $v$, the bandwidth of UAV-User communication denoted by
${B}$ is divided into a set of $\mathcal{L}$ subcarriers, each with bandwidth ${B}_1$ denoted by $l$, and the bandwidth of User-UAV communication indicated by $\tilde{B}$ that is divided into a set of $\mathcal{E}$ subcarriers, each with bandwidth $\tilde{B}_1$ denoted by $e$. Fig. \ref{u2u} shows the considered system model.
\vspace{-0.65cm}

\subsection{Communication Model}
\subsubsection{UAV to User Communication (DL Access Network)}
We consider a PD-NOMA technique for communication between UAVs and users, where the maximum transmit power of UAV $u$, $P^{\text{max}}_u$ is divided among its serving users over all subcarriers by adopting power allocation,  $\hat{p}_{uk}^l(t)$. In PD-NOMA, the transmitter can transmit signals of more than one user on each subcarrier using superposition coding (SC), and different users can receive its signal using successive interference cancellation
(SIC). To send information of user $k$ on subcarrier $l$ represented by $\hat{s}^l_{k}$, with unit power, i.e. $\mathbb{E}\{|\hat{s}^l_k|^2\}=1$, UAV $u$ sends the following signal on subcarrier $l$ at time slot $t$, $\zeta^l_u = \sum_{k\in\mathcal{K}}g^l_{uk}(t)\sqrt{\hat{p}_{uk}^l(t)}\hat{s}^l_{k}$, where $g^l_{uk}(t)$ is a binary variable with $g^l_{uk}(t)=1$ when user $k$ is assigned to UAV $u$ on subcarrier $l$ at time slot $t$ and otherwise $g^l_{uk}(t)=0$,
and we ensure that each user can only be allocated to only one UAV in DL as follows:
\begin{equation}\label{ps0}
\text{C4:} \left[ \sum_{l\in\mathcal{L}}g^l_{uk}(t)\right] \left[ \sum_{u'\in\mathcal{U},u'\neq u}\sum_{l\in\mathcal{L}}g^l_{u'k}(t)\right]=0, \forall u,k.
\end{equation}

Moreover, by the following constraint, we enforce that the transmit power of UAV $u$ over all subcarriers to its serving users does not exceed the power budget of UAV $u$, $P^{\text{max}}_u$:
\begin{equation}\label{eq-power-u}
\text{C5:} \sum_{k\in\mathcal{K}}\sum_{l\in \mathcal{L}}g^l_{uk}(t)\hat{p}_{uk}^l(t) \leq P^{\text{max}}_u.
\end{equation}
UAV $u$ sends signal to user $k$ on subcarrier $l$ through a channel with the channel power gain $\bar{\hat{h}}^l_{uk}(t)$, which consists of line of sight (LoS) and non-line of sight (NLoS) links as follows \cite{4}:
\begin{equation}\label{channel}
\bar{\hat{h}}^l_{uk}(t) = 
\begin{cases}
\hat{h}_{uk}^{l,\text{LoS}}(t)=\left( \frac{d_{uk}(t)4\pi f^l_c}{c} \right) ^{-\kappa}, &\quad \text{LoS},\\
\hat{h}_{uk}^{l,\text{NLoS}}(t)=\xi \left( \frac{d_{uk}(t)4\pi f^l_c}{c} \right) ^{-\kappa}, &\quad  \text{NLoS},\\ 
\end{cases}
\end{equation}
where $d_{uk}(t)$ is the distance between UAV $u$ and user $k$ at time slot $t$, $\kappa$ is the path loss component, $\xi$ is the additional path loss component for the NLoS link, $f^l_c$ denotes the carrier frequency of subcarrier $l$, and $c$ is the light speed. The probability of LoS link, i.e., $\text{Pr}_{\text{LoS}}$, and the probability of NLoS link, i.e, $\text{Pr}_{\text{NLoS}}$, are obtained as follows \cite{s2}:
\begin{equation}\label{eq-4}
\begin{multlined}
\text{Pr}_{\text{LoS}}= \text{Pr}\big(
\hat{h}_{uk}^{l,\text{LoS}}(t)\big)= \frac{1}{1+\beta_1 \exp\left(-\beta_2\left( \frac{180}{\pi}\arctan\left( \frac{
		\hat{h}_{uk}^{l,\text{LoS}}(t)}{d_{uk}(t)}\right) -\beta_1\right)\right)},
\end{multlined}
\end{equation}
\begin{equation}\label{eq-5}
\begin{multlined}
\text{Pr}_{\text{NLoS}} =\text{Pr}\big(\hat{h}_{uk}^{l,\text{NLoS}}(t)\big)= 1-\text{Pr}\big( \hat{h}_{uk}^{l,\text{LoS}}(t)\big),
\end{multlined}
\end{equation}
where $\beta_1$ and $\beta_2$ are related to the operation environment \cite{EN}. Therefore, the channel power gain is obtained as:
\begin{equation}\label{eq-6}
\bar{\hat{h}}^l_{uk}(t) = \text{Pr}_{\text{LoS}}~ \hat{h}_{uk}^{l,\text{LoS}}(t)+\text{Pr}_{\text{NLoS}}~ \hat{h}_{uk}^{l,\text{NLoS}}(t).
\end{equation}
The received signal at user $k$ on subcarrier $l$ assigned to UAV $u$ contains the signal that its UAV and other UAVs send on subcarrier $l$:
\begin{equation}
\psi^l_{uk}=\bar{\hat{h}}^l_{uk}(t)\zeta_u^l(t)+{\sum_{u'\in \mathcal{U}, u'\neq u}{g^{l}_{u'k}(t)\bar{\hat{h}}^l_{u'k}(t)}\zeta_{u'}^l(t)}+n^l_k.
\end{equation}\vspace{-.15cm}
From the PD-NOMA concept, to utilize 
SIC for separating different user's signals, an ordering among users served by each UAV should be achieved. The channel power gain to interference and noise ratio (CINR) is considered as the criterion for ordering users of an UAV\footnote{For SIC decoding, ordering is an important issue for improving the PD-NOMA performance. The optimal ordering can be obtained by solving the optimization problem to fine the optimal ordering variables. However, this approach has high computational complexity, so many works use CINR based ordering which is suitable for multi-UAV system with inter UAVs interferences \cite{moltafet2018optimal}.}, thus CINR of user $k$ assigned to UAV $u$ can be computed as below \cite{moltafet2018optimal}:
\begin{equation}
CINR_{uk}^l(t)= \frac{\bar{\hat{h}}^l_{uk}(t)}{\sum_{u'\in \mathcal{U}, u'\neq u}{\bar{\hat{h}}^l_{u'k}(t)}+B_1N_0},
\end{equation}
where $B_1$ denotes the bandwidth of each subcarrier, and $N_0$ is the power spectral density of noise on subcarrier $l$. The users are ordered in the decreasing order from the strongest user, i.e., the user with the highest CINR to the weakest user, i.e., the user with the lowest CINR for SIC decoding at the receiver sides. By the ordering of users, each user can remove the signal of other users which have lower CINR than its signal, and consider the signal of others which have higher CINR as noise. The instantaneous signal to interference and noise ratio (SINR) at user $k$ on subcarrier $l$ which is assigned to UAV $u$ is computed as follows:
\begin{equation}
\text{SINR}_{uk}^{l}(k,t) = \frac{g^l_{uk}(t)\bar{\hat{h}}^l_{uk}(t){\hat{p}_{uk}^l(t)}}{\bar{\hat{h}}^l_{uk}(t)\sum_{k'\in \mathcal{K}_u,CINR_{uk'}^l>CINR_{uk}^l}{g^l_{uk'}(t)\hat{p}_{uk'}^l}(t)+I_{uk}^l(t)+B_1N_0}
\end{equation}
where 
$I_{uk}^l(t)$ is the inter UAV interference on user $k$ which is assigned to UAV $u$ and is computed as $I_{uk}^l(t)= \sum_{u'\in\mathcal{U},u'\neq u}g^l_{u'k}(t)p_{u'k}^l\bar{\hat{h}}^l_{u'k}(t)$. The following constraint arises to perform the SIC successfully:
\begin{equation}\label{eq-order}
\text{C6:}~ \text{SINR}_{uk}^l(i,t) - \text{SINR}_{uk}^l(k,t)\geq 0, CINR_{uk}^l(i,t)>CINR_{uk}^l(k,t),\forall u\in\mathcal{U}, k\in\mathcal{K}, l\in\mathcal{L}.
\end{equation}
where $\text{SINR}_{uk}^l(i,t)$ denotes the SINR of user $k$ on subcarrier $l$ at UAV
$u$ at user $i$ \cite{tse2005fundamentals}. Based on this constraint, the SINR of worse user (with lower CINR) at the better users (with higher CINR) should be more than
its own SINR. The data rate for user $k$ of UAV $u$ on subcarrier $l$ at time slot $t$ with bandwidth $B_1$ is obtained as follows:
\begin{equation}\label{e1}
r_{uk}^l(t) =B_1\log_2 \big(1 + \text{SINR}_{uk}^l(t)\big) , \forall k \in \mathcal{K}, u \in \mathcal{U}.
\end{equation}
\subsubsection{User to UAV Communication (UL Access Network)}
For the user to UAV communication, we consider UL PD-NOMA scheme where multiple users transmit their signals to the UAVs. The received signal at UAV $u$ on subcarrier $e$ is given by
\begin{equation}
\varphi^e_u=\sum_{k\in\mathcal{K}_u}\tilde{g}^e_{ku}(t)\tilde{h}^e_{ku}(t)\sqrt{\tilde{p}_{ku}^e(t)}\bar{s}^e_{k}+\sum_{u'\in\mathcal{U},u'\neq u}\sum_{k\in\mathcal{K}_{u'}}\tilde{g}^e_{ku}(t)\tilde{h}^e_{ku}(t)\sqrt{\tilde{p}_{ku'}^e(t)}\bar{s}^e_{k}+\tilde{n}^e_u.
\end{equation}
Similar to DL access network, we define the parameters for UL access network on different subcarrier $e$, $\tilde{p}_{ku}^e(t)$ denotes the transmit power of user $k$ assigned to UAV $u$ on subcarrier $e$, $\bar{s}^e_{k}$ denotes the information of user $k$ transmitted on subcarrier $e$, with $\mathbb{E}\{|\bar{s}^e_{k}|^2\}=1$, $\tilde{n}^e_k$ is the Gaussian noise with power spectral density ${N}_0$, $\tilde{g}^e_{ku}(t)$ is a binary variable with $\tilde{g}^e_{ku}(t)=1$ when user $k$ is assigned to UAV $u$ on subcarrier $e$ at time slot $t$ in the UL and otherwise $\tilde{g}^e_{ku}(t)=0$, and we ensure that each user can only be allocated to one UAV in UL as follows:
\begin{equation}\label{ps2}
\text{C7:} \left[ \sum_{e\in\mathcal{E}}\tilde{g}^e_{uk}(t)\right] \left[ \sum_{u'\in\mathcal{U},u'\neq u}\sum_{e\in\mathcal{E}}\tilde{g}^e_{u'k}(t)\right]=0, \forall e,k.
\end{equation}
Based on (\ref{ps0}) and (\ref{ps2}), each user can be allocated to different UAVs separately in DL and UL transmission. In the following constraint, we enforce that the transmit power of user $k$ over all subcarriers to its assigned UAV does not exceed the power budget of user $k$, $\tilde{P}^{\text{max}}_k$:
\begin{equation}\label{eq-power-k}
\text{C8:} \sum_{u\in\mathcal{U}}\sum_{e\in\mathcal{E}}\tilde{g}^{e}_{ku}(t)\tilde{p}^e_{ku}(t)\leq \tilde{P}^{\text{max}}_k.
\end{equation}
User $k$ sends signal to  UAV $u$ on subcarrier $e$ through a channel with the channel power gain $\tilde{h}^e_{ku}(t)$.
The similar equations to (\ref{eq-4}), (\ref{eq-5}), and (\ref{eq-6}) can be used to determine the UL channel gains, $\tilde{h}^e_{ku}(t)$. 
By the ordering of users based on their channel gain, $\tilde{h}^e_{k'u}>\tilde{h}^e_{ku},\forall k', k \in \mathcal{K}_u$, the UAVs perform SIC to separate the signals of different users. The UAVs first decode the strongest user's signal by considering the signals of other users as noise, then after subtracting it from the received signal, decodes the next user's signal from the remaining, and so on. The instantaneous SINR of user $k$ on subcarrier $e$ at UAV $u$ is computed as follows:
\begin{equation}
\tilde{\text{SINR}}_{ku}^{e}(t) = \frac{\tilde{h}^e_{ku}(t){\tilde{p}^e_{ku}(t)}}{\tilde{h}^e_{ku}(t)\sum_{k'\in \mathcal{K}_u, \tilde{h}^e_{k'u}(t)<\tilde{h}^e_{ku}(t)}{\tilde{g}^e_{ku}(t)\tilde{p}^e_{k'u}(t)}+\tilde{I}_{u}^e(t)+\tilde{B}_1N_0},
\end{equation}
where $\tilde{I}_{u}^e(t)$ is the inter-user interference at UAV $u$ on subcarrier $e$ which is computed as below:
\begin{equation}
\tilde{I}_{u}^e(t)= \sum_{u'\in\mathcal{U},u'\neq u}\sum_{k\in\mathcal{K}_{u'}}
\tilde{g}^e_{ku'}(t)\bar{h}^e_{ku}(t)\tilde{p}_{ku'}^e(t).
\end{equation}
The received data rate from user $k$ on UAV $u$ on subcarrier $e$ with bandwidth $\tilde{B_1}$ at time slot $t$ is obtained as follows:
\begin{equation}\label{e1}
\tilde{r}^e_{ku}(t) =\tilde{B}_1 \log_2 \big(1 + \tilde{\text{SINR}}_{ku}^{e}(t)\big) , \forall k \in \mathcal{K}, u \in \mathcal{U}.
\end{equation}
\subsection{UAV to UAV Communication}\vspace{-0.15cm}
We consider a binary variable to show assignment of subcarriers as $\sigma_{uu'}^v(t)$, which is equal to 1 when subcarrier $v$ is utilized for communication between UAV $u$ to UAV $u'$, otherwise, 0. Subcarrier $v$ is chosen from subcarrier set $\mathcal{V}$ with size $V$. 
The data rate between two UAVs can be calculated as below:
\begin{equation}\label{w1}
\varpi^v_{uu'} = \hat{B}_1\log_2\left( 1 + \frac{\sigma_{uu'}^v(t)p_{uu'}^{v}(t)h_{uu'}^{v}(t)}{\hat{B}_1N_0}\right) ,
\end{equation}
where $p_{uu'}^v(t)$ is the transmit power over link UAV $u$ to UAV $u'$ on subcarrier $v$, and $h^v_{uu'}(t)$ is the channel power gain between UAVs $u$ and $u'$ computed as (\ref{channel}) and just contains the LoS link \cite{goddemeier2015investigation}. To satisfy orthogonality among subcarriers, the following constraint must be met:
\begin{align}\label{eq--1}
\text{C9:} \sum_{v\in\mathcal{V}}\sum_{u\in\mathcal{U}}\sum_{u'\in\mathcal{U}}\sigma_{uu'}^v(t)\leq 1.
\end{align}
\subsection{Network Function Virtualization}{We}
assume a set of communication service $\mathcal{O}_{\text{Total}} =\bigcup_{i=1}^{I} O_i$ where $I$ is the number of services. Service $O_i$ is described as $O_i = \{k^{\text{S}}_i,k^{\text{D}}_i, \mathcal{F}_i, b_i, T_i,\uptau_i \}$. Each service flow $i$, denoted by the pair of source
user and destination user $(k^{\text{S}}_i,k^{\text{D}}_i)$. $\mathcal{F}_i = \{ f_{i,1},\dots, f_{ij},\dots, f_{iJ} \}$ is a set of $J$ VNFs, where $f_{ij}$ denotes function $j$ of service $O_i$ and is assumed to be placed over UAVs. $b_i$ is the required bit rate of service $O_i$, $T_i$ denotes the time duration of service $O_i$, and $\uptau_i$ denotes the maximum delay tolerated by service $O_i$. Because of the limited resources, each UAV can run a few number of functions. Thus, the functions should be optimally placed by resource manager. In each time slot, we consider a binary function assignment variable $x_{uk}^{f_{ij}}(t)$ where $x_{uk}^{f_{ij}}(t) = 1$ means that function $f_{ij}$ is placed in UAV $u$ at time slot $t$ for user $k$. Note that in some cases, there is no direct links between two UAVs, therefore, some UAVs are utilized as relays which just receive the data and send to next nodes. We define a binary variable $y_{uk}^{O_i}(t)$ to represent if UAV $u$ act as a relay for $O_i$ service route of user $k$, $y_{uk}^{O_i}(t)=1$ and otherwise $y_{uk}^{O_i}(t)=0$. In addition, we define two binary variables $\chi_{unk}^{f_{ij}}(t)$ and $\mathcal{Y}_{unk}^{O_i}(t)$, to represent the path chain of each service. $\chi_{unk}^{f_{ij}}(t)$ is equal to 1 when function $f_{ij}$ of user $k$ is run in UAV $u$ and is sent to node $n$, $\mathcal{Y}_{unk}^{O_i}(t)$ is equal to 1 when node $u$ only relays the traffic of service $O_i$ of user $k$ toward node $n$, otherwise they are equal to 0. To ensure that each function can only placed in one node, we consider the following constraint:

\begin{equation}\label{ps}
\text{C10:} \sum_{u\in\mathcal{U}}x_{uk}^{f_{ij}} = 1, \forall i,j,k.
\end{equation}

In addition, with the considered flow conservation constraint, we ensure that for a requested service, a traffic leaves its source node $n$ and eventually arrives at its destination node $n'$ \cite{8075930} as follows:
\begin{equation}\label{ps1}
\text{C11:} \sum_{\acute{n}\in\mathcal{U}}(\chi_{n\acute{n}k}^{f_{ij}}+\mathcal{Y}_{n\acute{n}k}^{O_{i}})-\sum_{\acute{n}\in\mathcal{U}}(\chi_{\acute{n}nk}^{f_{ij'}}+\mathcal{Y}_{\acute{n}nk}^{O_{i}})=\begin{cases}
1&n=k^{\text{S}}_i,\\-1 & n=k^{\text{D}}_i\\
0 &\text{otherwise},
\end{cases},\forall i, n, k\in \mathcal{K},
\end{equation}

\begin{figure}
	\center
	\includegraphics[scale = .13]{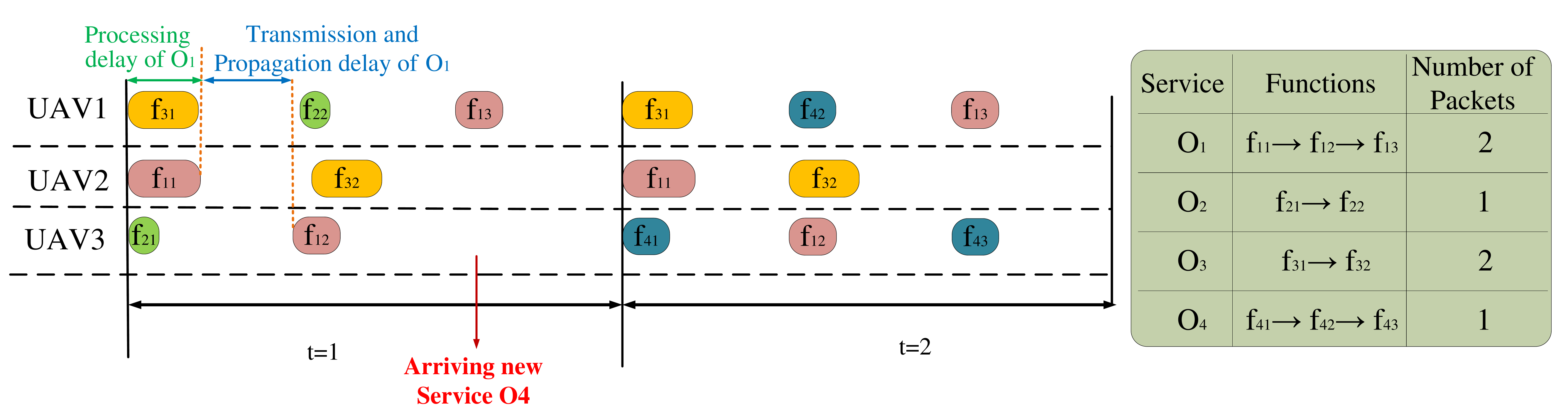}
	\center
	\caption{An example of procedure for management and scheduling of 3 UAVs for two time slots.}\label{fs1}
\end{figure}

The new requested services during a time slot are considered at the start of next time slot. An example of the procedure for management of 3 UAVs and two time slots is shown in Fig. \ref{fs1}. As it can be seen in Fig. \ref{fs1}, after processing a function, it is sent to the next node, which causes processing, transmission, and propagation delay.
The propagation delay depends on the distance between two communication nodes. The propagation delay between nodes $u$ and $u'$ for VNF $j$ of service $O_i$ of user $k$ at time slot $t$ is equal to
		\begin{subequations}
			\begin{align}
			\tau^{\text{PD},j}_{{iuu' k}}(t) =&\left( x_{uk}^{f_{ij}}(t) + y_{uk}^{O_i}(t)\right) \left( \mathcal{Y}_{uu'k}^{O_{i}}(t) + \chi_{uu'k}^{f_{ij}}(t)\right)  d_{uu'}(t)/c.
			\\
			& \label{consb} x_{uk}^{f_{ij}}(t) + y_{uk}^{O_i}(t) \leqslant 1,
			\\
			&\label{consc} \mathcal{Y}_{uu'k}^{O_{i}}(t) + \chi_{uu'k}^{f_{ij}}(t)\leqslant 1,
			\end{align}
		\end{subequations}
		and \eqref{consb} and \eqref{consc} ensure that each UAV just plays a roll as a function executor or relay.
The transmission delay of VNF $j$ of service $O_i$ of user $k$ depends on the assigned capacity of the utilized link capacity:
\begin{equation}
\tau^{\text{TD},j}_{{iuu'k}}(t) =\left( x_{uk}^{f_{ij}}(t) + y_{uk}^{O_i}(t)\right)
\left( \mathcal{Y}_{uu'k}^{O_{i}}(t) + \chi_{uu'k}^{f_{ij}}(t)\right) \frac{{b_i}(t)}{\varpi_{uu'}}, 
\end{equation}
where $\varpi_{uu'}$ is the total data rate capacity between two connected UAV $u$ and UAV $u'$. The processing delay depends on the function which should be executed in a node and the resource available in the node. We consider a processing delay of VNF $j$ of service $O_i$ of user $k$ at time slot $t$ at node $u$ as $\tau^{\text{PR},j}_{iuk}(t)$ which is computed as follows:
\begin{equation}
\tau^{\text{PR},j}_{iuk}(t)  = x_{uk}^{f_{ij}}(t)b_ic_{o}/\mathcal{C}_u,
\end{equation}
where $c_{o}$ is the required CPU cycle to process one bit. $\mathcal{C}_u$ is the total CPU cycle at UAV $u$. The utilized CPU cycle in each UAV should not exceed the available CPU cycle at each time slot, which is ensured as:
\begin{equation}\label{cpr}
\text{C12:} \sum_{k\in\mathcal{K}}\sum_{i\in\mathcal{I}}x_{uk}^{f_{ij}}(t)c_{o}b_i\leq \mathcal{C}_u, \forall u\in\mathcal{U}.
\end{equation}

The utilized bandwidth in each link should not exceed the available bandwidth at each time slot, which is ensured as:
\begin{equation}\label{BW}
\text{C13:} \sum_{k\in\mathcal{K}}\sum_{i\in\mathcal{I}}\left( \mathcal{Y}_{uu'k}^{O_{i}}(t) + \chi_{uu'k}^{f_{ij}}(t)\right)b_i\leq \sum_{v\in\mathcal{V}}\sigma_{uu'}^v(t)\varpi^v_{uu'}, \forall u,u'\in\mathcal{U}.
\end{equation}

\subsection{Dynamic VNF Migration}
Since the users and nodes are mobile in our proposed system and we consider time-varying traffic, dynamic VNF migration must be applied to existing VNF instances in order to meet QoS requirement of the newly arrived requests as well as providing continuous service for the existing users. In addition, the proposed dynamic VNF migration must have minimum impact on the QoS of the existing users. We formulate the dynamic VNF migration problem as a time-slotted model, where the position of UAVs and users are fixed at each time slot and changes from the current time slot to the next one. Note that a migration decision is made at the beginning of each time slot. We define the binary decision variable $m^{f_{ij}}_{\hat{u}\check{u}k}(t)$, with $m^{f_{ij}}_{\hat{u}\check{u}k}(t)=1$ if VNF $j$ of service $O_i$ of user $k$ assigned to UAV $\hat{u}$ at time slot $t-1$ migrates to UAV $\check{u}$ at time slot $t$, and $m^{f_{ij}}_{\hat{u}\check{u}k}(t)=0$ otherwise. Note that there is a relationship between $x^{f_{ij}}_{\hat{u}}(t-1)$,  $x^{f_{ij}}_{\check{u}}(t)$, and $m^{f_{ij}}_{\hat{u}\check{u}k}(t)$, given by $m^{f_{ij}}_{\hat{u}\check{u}k}(t)=x^{f_{ij}}_{\hat{u}}(t-1)x^{f_{ij}}_{\check{u}}(t)$. The flow conservation constraint also must be satisfied for each migration:
\begin{equation}\label{ps1-1}
\text{C14:} \sum_{\acute{u}\in\mathcal{U}}m_{u\acute{u}k}^{f_{ij}}(t)\chi_{u\acute{u}k}^{O_{i}}(t)-\sum_{\acute{u}\in\mathcal{U}}m_{\acute{u}uk}^{f_{ij'}}(t)\chi_{\acute{u}uk}^{O_{i}}(t)=\begin{cases}
1&,u=\hat{u},\\-1 & ,u=\check{u}\\
0 &,\text{otherwise},
\end{cases},\forall i, k\in \mathcal{K}.
\end{equation}
%
Migration of VNF $j$ of service $O_i$ of user $k$ from its current node $u'$ to its destination node $u$ may change the end-to-end delay of this service because the traffic of this service must be re-routed through different paths and VNF $f_{ij}$ must be executed in the new node with different computing capacity. Therefore, any VNF migration occurs only if it does not degrade the quality of corresponding service. The migration delay from UAV $u'$ to UAV $u$ for VNF $j$ of service $O_i$ of user $k$ at time slot $t$ can be formulated as:
\begin{equation}
\tau^{\text{MG},j}_{iuu'k}(t)=m_{u\acute{u}k}^{f_{ij}}(t)\frac{\alpha^{f_{ij}}}{\varpi_{uu'}}, 
\end{equation} 
where $\alpha^{f_{ij}}$ is information of VNF $j$ of service $O_i$ which must send from its source to its destination during migration process. 
The whole delay can be written as follows:
\begin{equation}
D(t) =
\sum_{n\in\mathcal{N}}\sum_{u\in\mathcal{U}}\sum_{k\in\mathcal{K}}\sum_{i\in\mathcal{I}}\sum_{j\in\mathcal{F}_i}
\bigg( \ \tau^{\text{PR},j}_{iuk}(t)+  \tau^{\text{PD},j}_{{iun k}}(t) + \tau^{\text{TD},j}_{{iun k}}(t)+\tau^{\text{MG},j}_{iunk}(t)\bigg).
\end{equation}
For each function of service $O_i$, we suppose that successful migration can happen when the destination VNF node has communication with the source VNF node and has enough resources to process the migrated function.
To ensure the QoS, the delay on service $O_i$ of user $k$ must
be smaller than the delay threshold $\tau_i$ as follows 
\begin{equation}\label{eq-delay}
	\text{C15:}
	\begin{cases}
 \sum_{n\in\mathcal{N}}\sum_{u\in\mathcal{U}}\sum_{j\in\mathcal{F}_i}\bigg( \ \tau^{\text{PR},j}_{iuk}(t)+     \tau^{\text{PD},j}_{{iunk}}(t) + \tau^{\text{TD},j}_{{iunk}}(t) +\tau^{\text{MG},j}_{iunk}(t)\bigg)\leq \tau_i,~~~~~~m^{f_{ij}}_{uu'k} = 1\\
\sum_{n\in\mathcal{N}}\sum_{u\in\mathcal{U}}\sum_{j\in\mathcal{F}_i}\bigg( \ \tau^{\text{PR},j}_{iuk}(t)+     \tau^{\text{PD},j}_{{iunk}}(t) + \tau^{\text{TD},j}_{{iunk}}(t)\bigg)\leq \hat\tau_i, ~~~~~~~~~~~~~~~~~~~\text{Otherwise}.
	\end{cases}
\end{equation} 
Constraint (\ref{eq-delay}) is considered when there is no VNF migration, the upper bound delay is decreased to $\hat\tau_i$ in order to prevent wasting of the resource usage. In other words, we assume that when there is no VNF migration, the sum of propagation, transmission, and processing delay is always lower than $\tau_i$.
\section{Problem Statement}
The optimization problem is formulated as below, in which we jointly maximize EE and minimize services' delay by finding optimal transmit power, subcarrier assignment, UAVs' location, service function chain path, function placement and scheduling subject to network and resource constraints.

\begin{figure}
\center
\includegraphics[scale = 0.2]{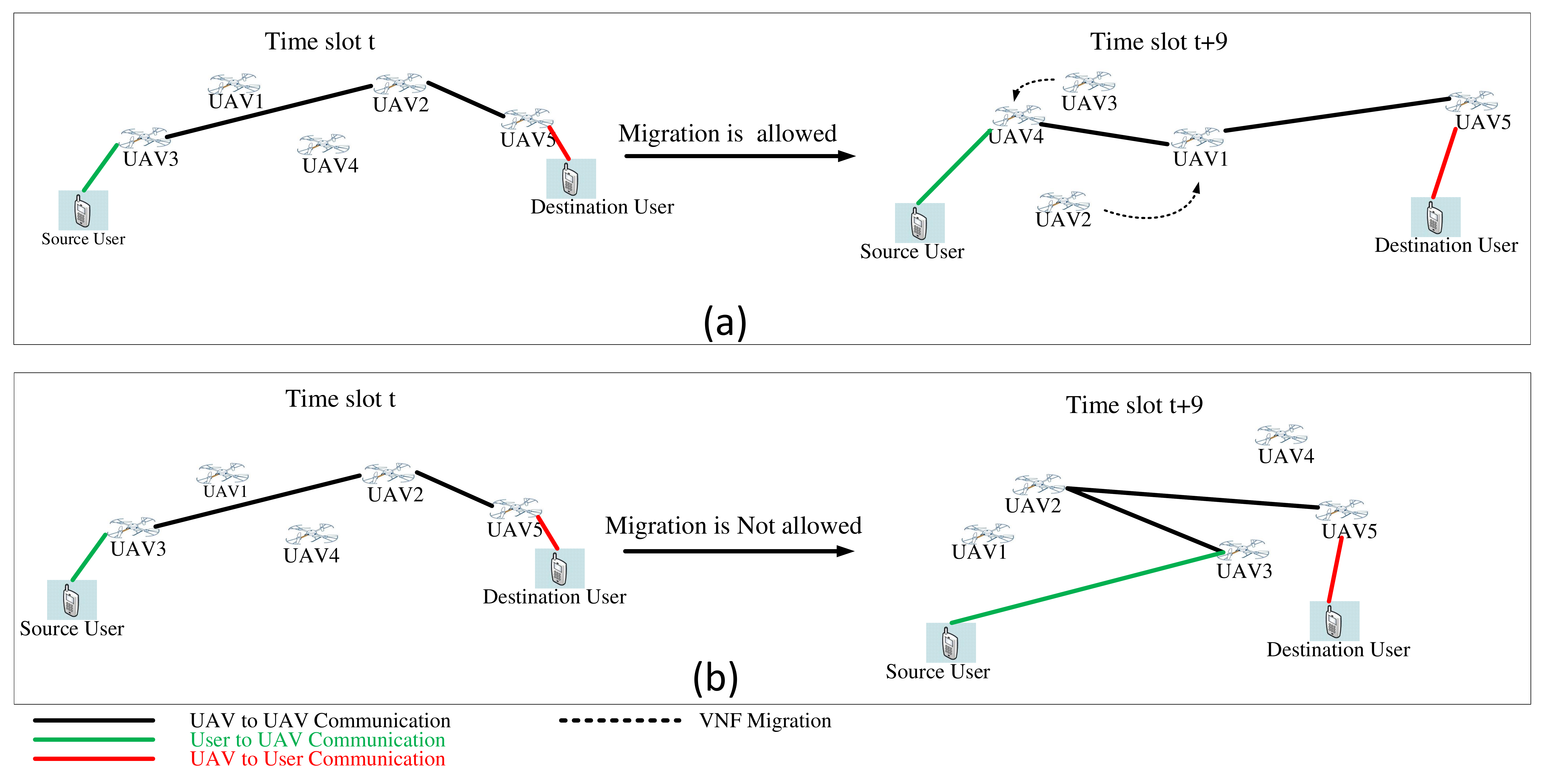}
\center
\caption{VNF placement, a)-With migration, and b)-Without migration}\label{Migration_Scenario}
\end{figure}

Defining EE as the ratio of the network sum-rate over the total consumed power, we can write the EE as follows \cite{EE}:
\begin{equation}\label{EE_eq}
E(t) = \frac{\sum_{u\in\mathcal{U}}\sum_{k\in\mathcal{K}_u}\sum_{l\in\mathcal{L}}g^l_{uk}(t)r^l_{uk}(t)+\sum_{k\in\mathcal{K}_u}\sum_{u\in\mathcal{U}}\sum_{e\in\mathcal{E}}\tilde{g}^e_{ku}(t)\tilde{r}^{e}_{ku}(t)}{\sum_{u\in\mathcal{U}}\sum_{k\in\mathcal{K}_u}\sum_{l\in\mathcal{L}}g^l_{uk}(t)p^l_{uk}(t)+\sum_{k\in\mathcal{K}_u}\sum_{u\in\mathcal{U}}\sum_{e\in\mathcal{E}}\tilde{g}^e_{ku}(t)\tilde{p}^e_{ku}(t)+P_{cr}(t)},
\end{equation}
where $P_{cr}(t)$ is the operation power consumption and is a function of UAV's movement and the time of activity \cite{19}, which can be hence defined as
\begin{equation}
P_{cr}(t) = \sum_{u\in\mathcal{U}}P_d{\xi}_u(t) + U P_c \iota,
\end{equation}
where $P_d$ is the power consumption per meter, $\xi_u$ is the amount of movement of UAV $u$ at time slot $t$, $P_{c}$ is the amount of constant power consumption of each UAV in one second.
\begin{equation}\label{opt1}
\begin{array}{rlllll}
\displaystyle {\min_{\boldsymbol{p},\hat{\boldsymbol{p}},\tilde{\boldsymbol{p}}, \check{\boldsymbol{q}}, \tilde{\boldsymbol{q}}, \boldsymbol{x},\boldsymbol{y},\boldsymbol{\chi},\boldsymbol{\mathcal{Y}},\boldsymbol{\sigma},\tilde{\boldsymbol{g}},\boldsymbol{g}}}& \multicolumn{1}{l}{\eta =\varrho_1E- \varrho_2 D,} \\
\textrm{Subject to:} &  \sum_{u\in\mathcal{U}}\sum_{l\in\mathcal{L}}\alpha^{O_i}_kg^l_{uk}(t)r_{uk}^l(t)\geq \alpha^{O_i}_k(t)b_i,\\
&\sum_{u\in\mathcal{U}}\sum_{e\in\mathcal{E}}\alpha^{O_i}_k\tilde{g}^e_{ku}(t)\tilde{r}^e_{ku}(t)\geq \alpha^{O_i}_k(t)b_i,\\
&\sum_{v\in\mathcal{V}}\sum_{u\in\mathcal{U}}\sum_{u'\in\mathcal{U}}\alpha^{O_i}_k(t)\sigma_{uu'}^v(t)\varpi^v_{uu'}\geq \alpha^{O_i}_k(t)b_i,\\
&\displaystyle  \tau^{\text{PR}}_{d^{f_{ij}}_{u k}}(t)  +  \tau^{\text{PD}}_{{iun k}}(t) + \tau^{\text{TD}}_{{iun k}}(t)+\tau^{\text{mg}}_{iunk}(t)\leq \iota, \forall k \in \mathcal{K},\\
& \text{C1}-\text{C15},
\end{array}
\end{equation}
where $\boldsymbol{p}=[p^v_{uu'}],\hat{\boldsymbol{p}}=[\hat{p}_{uk}^l(t)],\tilde{\boldsymbol{p}}=[\tilde{p}^e_{ku}], \check{\boldsymbol{q}}=[\check{q}_u], \tilde{\boldsymbol{q}}=[\tilde{q}_k], \boldsymbol{x}=[x_{uk}^{f_{ij}}],\boldsymbol{y}=[y_{uk}^{O_i}],\boldsymbol{\chi}=[\chi_{unk}^{f_{ij}}],\boldsymbol{\mathcal{Y}}=[\mathcal{Y}_{unk}^{O_i}],\boldsymbol{\sigma}=[\sigma_{uu'}^v],\tilde{\boldsymbol{g}}=[\tilde{g}^e_{ku}(t)],\boldsymbol{g}=[g^l_{uk}(t)]$, $\varrho_1$ and $\varrho_2$ are two coefficients to balance the EE and the whole average delay, $\alpha^{O_i}_k(t)$ is one if user $k$ requests service $O_i$, and otherwise is zero. The problem is solved at each time slot in two different scenarios. In the first scenario, we consider that the migration can happened for each VNF of each user while in the second scenario, there is no migration during the time duration of service for each user. For example, consider user $k_i^{S}$ requests service $O_i$ to the user $k_j^{D}$ with service duration $T_i=10$ (time slots). According to the first scenario, every 10 time slots, the VNF migration can be happen to guarantee the user QoS and minimize cost of the network, on the other hand, in the second scenario, there is no VNF migration until the end of 10th time slot. As shown in Fig. \ref{Migration_Scenario}, if migration is allowed (Fig. \ref{Migration_Scenario} (a)), then the VNF placement may change based on the user's requirements, user's mobility, and network cost, and if migration is not allowed (Fig. \ref{Migration_Scenario} (b)), then the VNF placement remains the same until the 10$th$ time slot.
%

\section{Proposed Algorithm}
The optimization problem in (\ref{opt1}) is non-convex and NP-hard. Thus, in order to solve it, we adopt a hybrid action DRL-based method for centralized (single-agent) and decentralized (multi-agent) scenarios. The optimization variables are UAV's trajectories, subcarrier assignment, transmit power allocation, and VNF placement and scheduling. The problem is solved at each time slot based on new locations of users and the requested services.
Our optimization problem can be formulated as MDP, where UAVs act as agents and the Internet cellular network consisting of users is considered as environment. At each time slot, the UAVs act in the Internet cellular environment and serving the user's demands which causes the environment to transitions to another state. Considering (\ref{opt1}) as MDP yields to solve that using DRL methods. The basic common RL method is Q-learning which suffers from limitation in discrete and low dimensional state and action spaces \cite{gr}. On the other hand, using neural network as function approximation for Q-learning, the DQN could be used to solve high dimensional continuous state spaces. However, DQN can only handle low dimensional discrete actions \cite{ql0}. To overcome the high dimensional and continuous action spaces, DDPG, a model-free off-policy reinforcement learning method \cite{ql0}, could be used. DDPG uses four networks as deterministic policy network (Actor), Q network (Critic), target policy network (target Actor) and target Q network (target Critic). The actor outputs continuous actions from continuous states directly instead of mapping the probability distribution across discrete action space. Using the selected actions and the states as input, the Critic network outputs the Q value in order to define how good or bad is those actions. Then the Actor updates the policy distribution in the direction suggested by the critic. DDPG reinforcement learning is robust and appropriate solution for the environment which the states and actions are continuous. However, in some situations, agents must perform both discrete and continuous actions simultaneously. In such situations, it’s convenient to use methods that can play with a combination of discrete and continuous actions \cite{hybridDC}. One of the typical approach to handle hybrid continuous and discrete actions is to convert some of the continuous output of actor network into discrete actions \cite{hybridDC}. Alternatively, by quantizing continuous actions of action space, we have fully discrete action space so that the methods like DQN could solve the problem \cite{quantizeDisc}. None of the mentioned methods are optimal for mixed continuous and discrete actions since mapping continuous actor outputs to discrete actions for DDPG suffers from slow learning while quantizing continuous actions in action space for DQN causes high oscillations in the converging steps \cite{hybridDC}. 

\subsection{Proposed Hierachical Hybrid Continuous and Discrete Actions}\label{HHCDA}
\subsubsection{Single Agent}
\begin{figure}[h]
	\begin{center}
	\includegraphics[scale = 0.15]{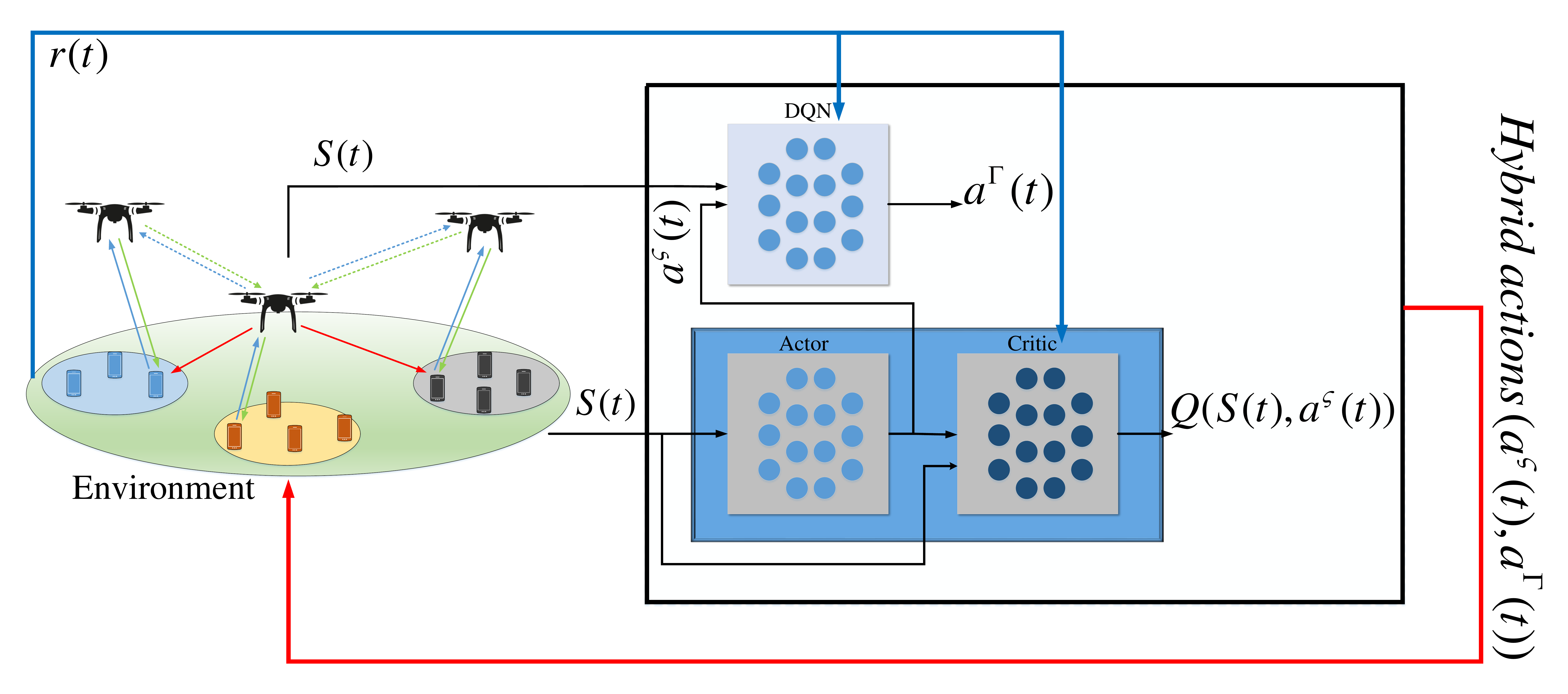}
	\caption{Hierachical Hybrid Continuous and Discrete Actions method.}\label{PDRL}
	\end{center}
\end{figure}

In this section, we introduce our proposed DRL method which can support both continuous and discrete actions simultaneously. Consider an agent with an action $a \in \mathcal{A}$ which is a combination of multiple continuous and discrete values as $a^{\varsigma}=(a_{1}^{\varsigma}, ... , a_{\xi}^{\varsigma})$ and $a^{\varGamma} = (a_{1}^{\varGamma}, ... , a_{\digamma}^{\varGamma})$, respectively, where $\digamma$ and $\xi$ denote the number of discrete and continuous actions.
Therefore each action includes a set of continuous and discrete values, i.e., $a=(a^{\varsigma}, a^{\varGamma})$. For the continuous actions, the DDPG method and for the discrete action, DQN is considered. It is worth mentioning that the action selection mechanism of our proposed DRL method is performing in two successive steps. At the first step, the continuous actions are selected from the actor network in DDPG module using the state of the environment as input, then at the second step, the discrete actions are selected from the DQN module using the continuous actions plus the state of the environment as input. Let $\theta$, $\omega$, and $\phi$ denote the parameters for actor, critic and DQN networks, respectively. In the first step, the actor selects the appropriate actions based on an approximation of a deterministic policy $\pi(a^{\varsigma}(t)|s(t);\theta)$, then the critic estimates the goodness of actions taken by the actor with $Q(s(t),a^{\varsigma}(t);\omega) = \mathbb{E}_{a^{\varsigma}(t)}\Big[R(t)|s(t),a^{\varsigma}(t)\Big]$, where the expectation taken over the action $a^{\varsigma}(t)$ is based on the deterministic policy $\pi$, i.e., $\pi(s(t)|a^{\varsigma}(t);\theta)$. Accordingly, at each training step $t$, the critic calculates the loss using the following equation:
\begin{equation}
	L(\omega)= \mathbb{E}_{s,a^{\varsigma},r,s'\sim \mathcal{B}} \Big[(y(t)-Q(s(t),a^{\varsigma}(t);\omega))^2\Big],
\end{equation}   
where $y(t) = r(t)+\gamma Q_{\omega'}(s(t+1),a^{\varsigma}(t+1))$, $\omega'$ is the parameter of the target critic network, and $\mathcal{B}$ is the replay buffer. The aim for DDPG module is to find a policy $\pi$ which gives the action $a^{\varsigma}(t)$ that maximizes the $Q_{\omega}(s(t),a^{\varsigma}(t))$. Therefore, by performing the gradient ascend of loss function with respect to $\theta$, we aim to solve the following objective:
\begin{equation}
	\max\limits_{\theta} \mathbb{E}_{s\sim \mathcal{B}} \Big[Q_{\omega}(s(t),a^{\varsigma}(t))\Big].
\end{equation}
After selecting the appropriate continuous actions at time slot $t$ as $a^{\varsigma}(t)$, the DQN module outputs the discrete actions using the $s_{DQN}(t)$ which is sum of state $s(t)$ and $a^{\varsigma}(t)$ as $s_{DQN}(t)=a^{\varsigma}(t)\cup s(t)$. Fig. \ref{PDRL} depicts our proposed HHCDA in single-agent mode.
For our HHCDA agent, we can define a tuple as $(\mathcal{S}, \mathcal{A}, \mathcal{R}, \mathcal{S}^\prime, \gamma)$ so that $\mathcal{S}$ is the agent's observation from the environment, $\mathcal{A}$ is all possible agent actions, $\mathcal{S}^\prime$ is the next state, $\mathcal{R}$ is the reward function and $\gamma$ is the discount factor which defines how the future reward is important. More precisely, we define the elements of each part of the tuple as follows:
\begin{itemize}
	\item{
		
		The state space $\mathcal{S}$: The state of the agent at time slot $t$ consists of all UAVs and users position as $\boldsymbol{\check{q}(t)}$ and $\boldsymbol{\tilde{q}(t)}$, respectively, as well as the arrived services of user $k$, as $O_k = \{k^{\text{S}}_k,k^{\text{D}}_k, \mathcal{F}_k, b_k, T_k,\uptau_k\}$ and the residual cpu resource of all UAVs as $\mathcal{C}_{U}$ and the residual bandwidth between UAVs as $\varpi_{uu'}$. So the state space for all agent is the same and defined as $\mathcal{S} =(\boldsymbol{\check{q}(t)}, \boldsymbol{\tilde{q}(t)},O_k,\mathcal{C}_{U}, \boldsymbol{\sigma})$.}
	
	\item{
		The action space $\mathcal{A}$: The action space consists of UAVs movement, $\boldsymbol{\Delta \check{q}_U}$, uplink power allocation, $\boldsymbol{\hat{p}(t)}$, downlink power allocation, $\boldsymbol{\tilde{p}(t)}$, VNF placement, $\boldsymbol{\chi(t)}$, VNF relay, $\boldsymbol{\mathcal{Y}(t)}$, uplink subcarier allocation, $\boldsymbol{\tilde{g}(t)}$, and downlink subcarrier allocation $\boldsymbol{{g}(t)}$.
		So, the action space can be defined as  ($\boldsymbol{\Delta \check{q}_U},\boldsymbol{\hat{p}(t)},\boldsymbol{\tilde{p}(t)},\boldsymbol{\chi(t)},\boldsymbol{\mathcal{Y}(t)},\boldsymbol{\tilde{g}(t)},\boldsymbol{{g}(t)}$).}
	
	\item{
		The reward function $\mathcal{R}$:
		The reward is a numerical value given to the agent at each time slot $t$ from the environment and shows how the agent performs in the environment and achieves its goal. The reward function for our proposed model can be defined as:
		\begin{align}\label{reward}
			\mathcal{R} =\varrho_1E- \varrho_2 D.
		\end{align} 
	}
\end{itemize}
\subsubsection{Multi Agent}
Considering the multi-agent reinforcement learning, we assume each UAV as an agent and each of them has DDPG and DQN for selecting the continuous and discrete actions individually. On the other hand, we consider a centralized critic for evaluating the actions of all UAVs in a centralized manner. Each UAV selects the action consisting of 17 variables as its movement towards $x$ and $y$ directions, transmit power and subcarrier allocation in both uplink and downlink transmission as well as VNF placement of 3 distinct users apart from their positions, distances, and scheduling. For instance, UAV1 supports three distinct users as user1, user2 and user3, and UAV2 supports other ones as user4, user5 and user6 and so on. It is worth mentioning that we consider that UAVs do not have full observation from the environment, they have limited information from the environment. Finally, for the reward function, the UAVs aim to maximize their team reward which is calculated as $\mathcal{R}$ based on (\ref{reward}).
In the proposed HHCDA method, the discrete actions are dependent on the continuous ones. In other words, the UAVs change their position in order to perform VNF placement. In this regard, the continuous actions related to UAVs trajectory design, as well as uplink and downlink transmissions powers ($a^{\varsigma}(t)$), will be used as input state for the DQN to output discrete actions ($a^{\Gamma}(t)$) as VNF placement and corresponding subcarriers for uplink and downlink transmissions.
	\begin{algorithm}[t]	
	\small
	\caption{The  multi-agent HHCDA method}	\label{MAHHCDA}
	\textbf{Input}: Initialize weights of actor, critic and DQN networks,  $\theta$, $\omega$, and $\phi$, with random value. \\
	\textbf{Set}: Put the value of main networks into the target networks, $\theta \rightarrow \hat{\theta}$ and $\phi \rightarrow \hat{\phi}$.  \\ 
	By considering the memory size of: $D_{\text{Mem}}$, mini-batch memory  size $D_{\text{Mini-Batch}}$, time slot number $T_d$, and terminate episode number $E$, perform following steps\\
	\For{ e = 1:E }{
		\For{$t_d$ = 1:$T_d$}{
			Observe $s(t)$ of environment \\
			\For{{$i \in$ (total UAVs)} }{
				Observe state $i$ from the environment\\
				Obtain the continuous actions $a^{\varsigma}_{i}(t)$ from $s_i(t)$ based on the policy $\pi$ from the actor network\\
				Obtain the discrete actions $a^{\varGamma}_i(t)$ from $s_i(t)+a^{\varsigma}_i(t)$ based on the policy $\pi_d$ from DQN network\\
				Execute the hybrid action $a^{\varsigma}_i(t)+a^{\varGamma}_i(t)$ for each UAV $i$  and obtain the team reward $r(t)$\\
				Environment transitions to new state $s(t+1)$\\
			}
			Store $(s(t), a^{\varsigma}(t), r(t), s(t+1))$ in expericence replay $D$ \\
			Store $(s_{DQN}(t), a^{\varGamma}(t), r(t), s_{DQN}(t+1))$ in expericence replay $D_{DQN}$ \\
		}
		Randomly sample from mini-batch of $D$\\
		Calculated following equation to update the critic\\
		$L(\omega)=\mathbb{E}_{D} ((y(t)_{i}-Q(s(t),a^{\varsigma}(t),\omega))^2)$\\
		$y_i(t)=r(t)+\gamma Q(s(t+1),a^{\varsigma}(t+1),\omega)$\\
		Calculated following equation to update the DQN\\
		$L(\phi)=\mathbb{E}_{D}((y^{d}_i(t)-Q(s_{DQN}(t), a^{\varGamma}(t);\phi))^2)$\\
		$y^{d}_i(t)=r(t)+\gamma \max \limits_{a^{\varGamma}(t+1)}Q(s_{DQN}(t+1), a^{\varGamma}(t+1);\phi)$\\
		Update the main networks using adam optimizer\\
		Update the wight of target networks by period $\tau$ \\		
	}
\end{algorithm}
\section{Computational Complexity Analysis}\label{ComputationalComplexity}
There are two main parts for the computational complexity of our proposed HHCDA DRL method: action selection complexity and training process complexity.

\subsection{Computational Complexity of Action Selection}
We consider a fully connected neural network with fixed numbers of hidden layers and fixed numbers of neurons in each hidden layer. The computational complexity of such networks is equal to sum of sizes of input and output. For each UAV, the sizes of the inputs of the actor, critic, and the DQN networks, i.e., the size of state and action spaces are $3U+2K+IJK+U+U(U-1)$, $3U+2K+IJK+U+U(U-1)+3U+5K$, and $3U+2K+IJK+U+U(U-1)+3U+2K$, respectively. Therefore, the computational complexity of UAV trajectory design, power allocation for all links, user association, subcarrier assignment, function placement, and changing route of services' traffic and estimation of the Q-function value for a state-action pair is $IJK$. All UAVs should estimate the Q-function values, thus the computational complexity of action selection is $\mathcal{O}(IJKU)$.

\subsection{Complexity of Training Process} 
The UAVs should calculate the Q-function values of all users. In accordance to previous section, this step has $\mathcal{O}(IJKUM)$ computational complexity where $M$ is the size of the training batch. Moreover, fully connected neural network with fixed number of hidden layers and neurons, the complexity of back-propagation algorithm is related to the product
of the input size and the output size. For each UAV, the sizes of the inputs of the actor, critic, and DQN networks are $3U+2K+IJK+U+U(U-1)$, $3U+2K+IJK+U+U(U-1)+3U+5K$, and $3U+2K+IJK+U+U(U-1)+3U+2K$, respectively. In addition, for each UAV, the sizes of the outputs of the actor, critic, and DQN networks are $3U+5K$, 1, and $2K$, respectively. Therefore, the computational complexity of the back-propagation algorithm is $\mathcal{O}(KU^3)$. Finally, the training process complexity for all UAVs is $\mathcal{O}(MKU^3)$. 

\section{Convergence Analysis}\label{Convergence Analysis}
For Q-learning algorithm, the Q-function can converge to the optimal Q-function as $t\rightarrow\infty$ with probability 1, if $\sum_{t=0}^{\infty}\alpha^t=\infty$ and $\sum_{t=0}^{\infty}(\alpha^t)^2<\infty$ are satisfied and $|r^t(s^t,a^t)|$ be bounded \cite{watkins1992q}. Since our proposed algorithm is an extended version of Q-learning algorithm, it can converge to the optimal Q-function as the mentioned condition are satisfied. To the fast convergence and train our neural network effectively, we utilize the inverse time decaying learning rate that uses the large learning rate in the first episodes in order to prevents the network from getting stuck in a bad local optimum trap near the initial point and uses the small learning rate in the last training epochs in order to converge to a good local optimum \cite{you2019does}. The convergence of our proposed algorithm is also analyzed through simulations in Section \ref{sect4}.

\section{Performance Evaluation}\label{sect4}

\begin{table}[ht]\label{table3}
	\small 
	\setlength{\tabcolsep}{8pt}
	\caption{Network Parameters} 
	\centering          
	\begin{tabular}{|l  | l|l|} 
		\hline     
		Parameter & Description& Value\\
		\hline
		$K$&\text{Number of users} & 6/9/12/15\\
		\hline
		$U$&\text{Number of UAVs} & 6\\
		\hline
		\textcolor{black}{H}& UAVs altitude &75 m \cite{wu20205g}\\
		\hline
		$W^{max}$& UAVs speed &10 m/s\\
		\hline
		& Maximum UAV CPU capacity &1 GHz (32 bits)\\
		\hline
		& Maximum UAV MEMORY capacity &2 GB\\
		\hline
		$V^{max}$& user mobility(Random walk) &0/3/6/9 km/h \\
		\hline
		$R^{\text{min}}$&Minimum required rate &5 Mbps\\
		\hline
		$\beta_1, \beta_2$& Environment parameters& 0.36, 0.21\\
		\hline
		$f_c$&Carrier frequency& 2 GHz \\
		\hline
		$P_{\text{c}}$&Constant power consumption per second& 0.01 Watts \cite{hu2020cooperative}\\
		\hline
		$P_{\text{d}}$& Power consumption per meter &0.05 Watts \cite{hu2020cooperative}\\
		\hline
		$\hat{p}_{uk}$& Maximum transmit power for UAV to users &5 Watts \\
		\hline
		$\tilde{p}_{ku}$& Maximum transmit power for UAV to UAV &5 Watts \\
		\hline
		$\iota$&Time slot interval &0.5 s\\
		\hline
		$\alpha$&Initial learning rate &0.001\\
		\hline
		$\gamma$&Discount factor &0.99\\
		\hline
		$\hat{B}, B, \tilde{B}$&Bandwidth for all 3 communications &5 MHz\\
		\hline
		$N_0$&Noise power spectral &-170 dBm/Hz\\
		\hline
		$\kappa$&Path loss exponent&3.5\\
		\hline
		$\varepsilon$& Epsilon decay &0.01 \\
		\hline
		--&Activation function& ReLU and tanh \cite{wu2020reducing}\\
		\hline
		--&Number of episodes& 3000\\
		\hline
		--&Number of iterations& 300000\\
		\hline
		--&Number of hidden layers&3 \\
		\hline
		--&Target network update frequency& 1000 \cite{wu2020reducing}\\
		\hline
		$M$&Batch size& 128\\
		\hline
		--&Number of neurons in each layer & 512\\
		\hline
	\end{tabular}%
\end{table}
	{In this section, the performance of our proposed model NFV enabled aerial network is analyzed using different DRL methods. Also, the comparison of our proposed single-agent HHCDA and Multi- Agent HHCDA (MAHHCDA) DRL solution with the state-of-the-art ones, i.e., DDPG and Multi-Agent DDPG (MADDPG) are provided accordingly. For the performance assessment, we consider a square-shaped area with the length of 1000 $m$ where there are 12 users that are randomly and uniformly distributed in the area \cite{99333}. We consider 6 UAVs which can move at 10 $m/s$ as maximum speed toward the $x$ and $y$ directions and have 75 $m$ altitude. Each user requests a service which is described with the destination user, number of functions, bit rate and the maximum tolerable delay. The simulation is conducted using Tensorflow \footnote{The source code and implementations are provided in https://ieee-dataport.org/documents/ai-based-and-mobility-aware-energy-efficient-resource-allocation-and-trajectory-design-nfv }. Other simulation parameters are provided in Table \ref{table3}. \\
	The trajectories of UAVs and users during training process are depicted in Fig. \ref{trajectories}. The starting position of each UAV is demonstrated by a square. After several iterations, as the UAVs trained, the movements of UAVs are limited to smaller area which is depicted with a circle as the final position of them. The comparison of the final position of each UAV with the users positions indicates that the agents (UAVs) are trained to near optimal positions.

	\begin{figure}
	\begin{center} 
		\subfigure[]{\label{UAV-trajectory}\includegraphics[width=5.5cm]{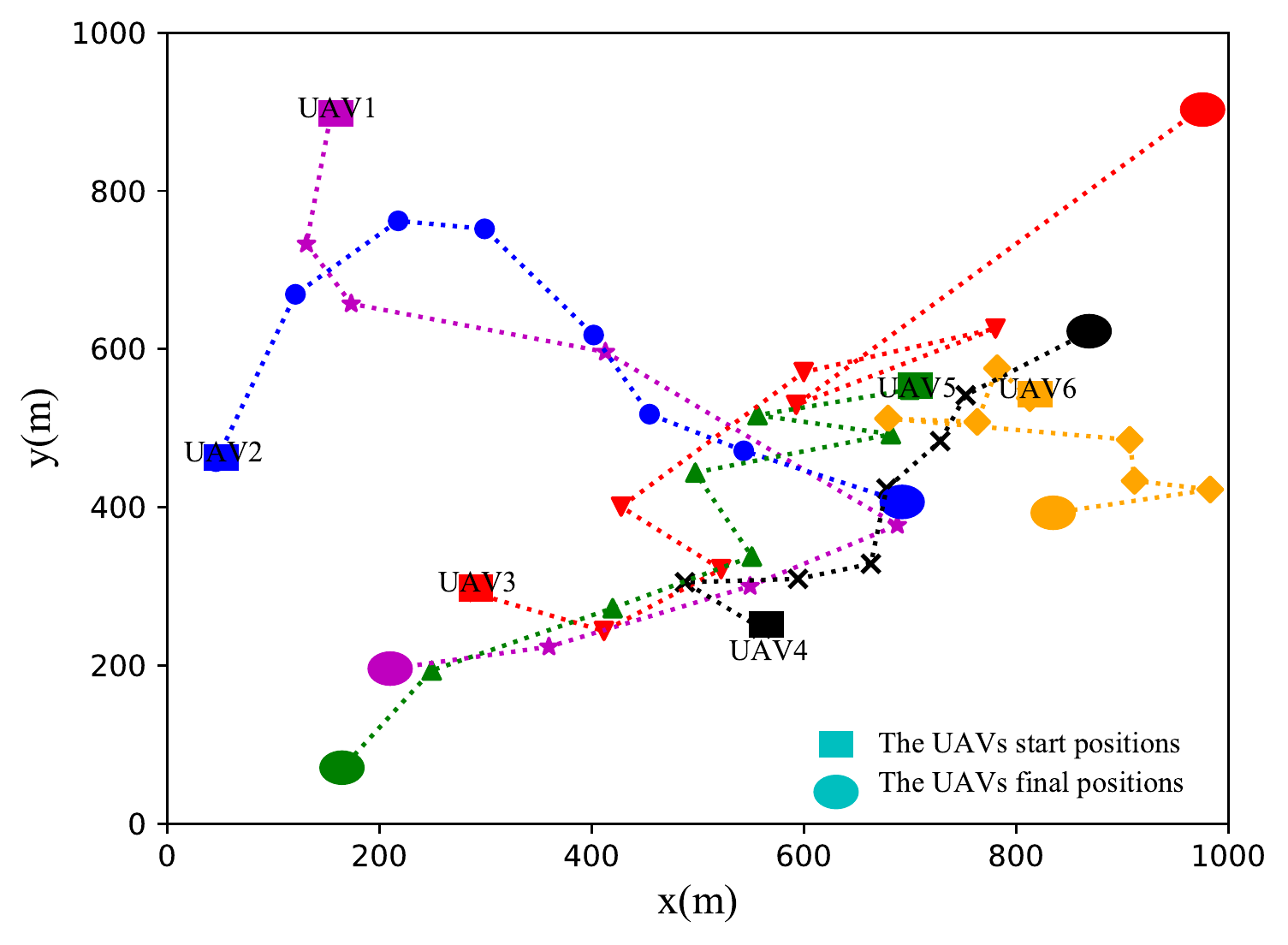}}
		\subfigure[]{\label{user-trajectory}\includegraphics[width=5.7cm]{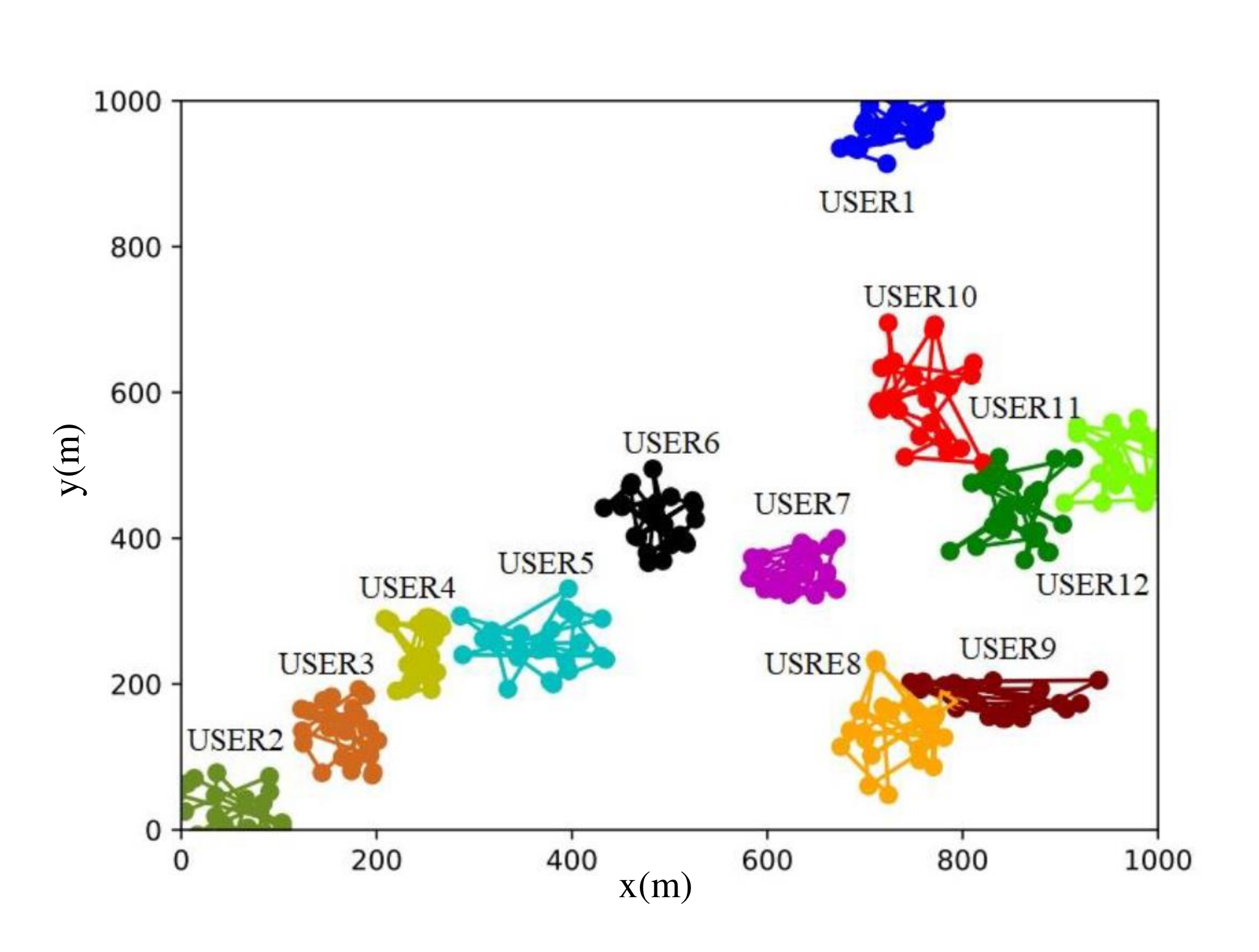}}
		\subfigure[]{\label{Rewards}\includegraphics[width=5.3cm]{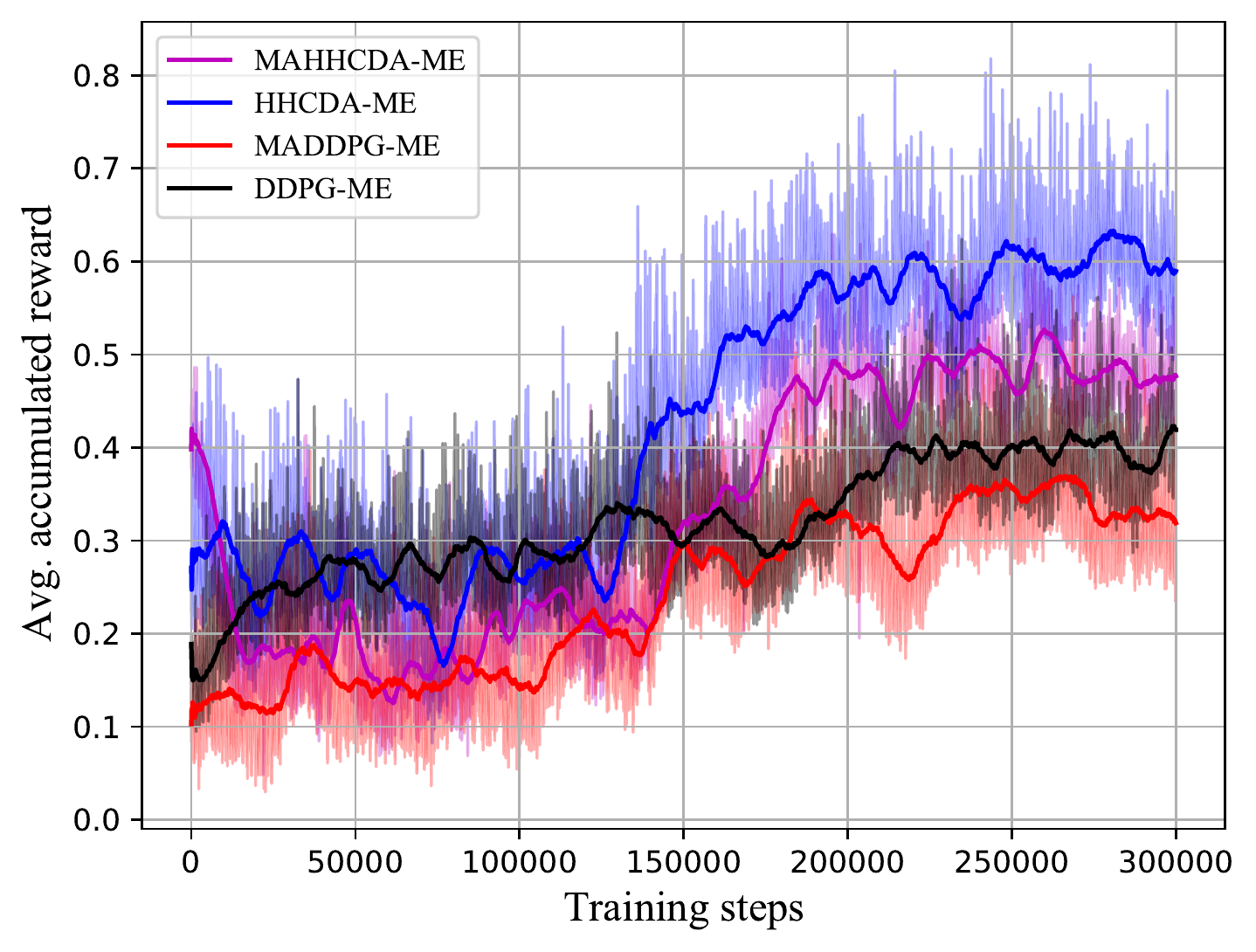}}
		\caption{(a) The UAVs trajectories during training process, (b) The users movement during training process, and (c) The average accumulated reward for different methods with VNF migration and considering 12 users with 3 km/h mobility} \label{trajectories}
	\end{center}
	\end{figure}

	Fig. \ref{trajectories} (c) shows the accumulated reward (pallid curves) with its average (solid curves) obtained by our proposed MAHHCDA Migration Enabled (MAHHCDA-ME) and HHCD-ME compared to state-of-the-art methods, MADDPG-ME and DDPG-ME, during the training process. Note that the "ME" prefix for all DRL methods is used for indicating that the migration is allowed for that method. As depicted, our proposed single-agent HHCDA-ME requires less training steps than other methods. On the other hand, MAHHCDA-ME converges to lower reward value with higher training steps compared with HHCDA-ME, because in our proposed MAHHCDA, the agents have limited communication with each other, i.e., the agents only know about other agents' tasks and their instantaneous positions. Whereas, the single-agent HHCDA has whole information about the environment and accordingly, it outperforms the multi-agent HHCDA. This is true for MADDPG and DDPG as well. \textcolor{black}{Although} two single-agent methods perform better than multi-agent ones, they have very high computational complexity and signaling overhead. It is worth mentioning that we can improve the multi-agent performance, even more than single-agent methods, by considering full communication between agents where all agents have full observation from the environment \cite{MA-signal-processing}.
	\\
	Fig. \ref{mobility_impacts} (a) and Fig. \ref{users_impacts} (a) show the RRR versus the mobility of users and  the number of users with 12 users having 3 km/h mobility, respectively. As demonstrated, increasing the mobility as well as the number of users will increase the RRR significantly. On the other hand, considering VNF migration, shown as methods with "\_ME" prefix in the figures, yields to decrease the RRR compared with other methods because the agents are allowed and trained to perform suitable VNF migration based on the mobility of users and their requests. For the mobility of users equal to 0 $\text{km/h}$, both MAHHCDA-ME and MAHHCDA, or HHCDA-ME and HHCDA, have similar RRR, but as the the mobility of users increases, the methods which are allowed to do VNF migration, outperforms other ones. Another result from Fig. \ref{mobility_impacts} (a) is that, with the RRR equal to 5\% and 12 users with 3 $\text{km/h}$ mobility, only HHCDA-ME supports the network and user's requirements. The average delay comparisons are provided in Fig. \ref{mobility_impacts} (b) and Fig. \ref{users_impacts} (b) for different DRL methods. Obviously, considering managed and proposed  VNF migration rather than random and unmanaged one for DRL methods, causes significant decrease in average delay. This is expected because when there is no VNF migration, as the users services chain established on the UAVs, they may change their positions at each time slot during the life time of their services, and cause to change their channel power gains and introduce more delay or interference to the network. It should be mentioned that we consider 100 ms as maximum tolerable average delay and all DRL methods can handle the network delay requirement. The comparisons of the EE are shown in the Fig.~\ref{mobility_impacts} (c) and Fig.~\ref{users_impacts} (c). As depicted, increasing the mobility of users as well as the number of users will decrease the EE since with higher mobility, the UAVs encounter with more out-dated channel power gains, resulting a remarkable decrease for EE. Although increasing the number of users will increase the data rate of the network in the first place, we consider the users uplink power as the consumed energy in the denomerator of the EE expression in (\ref{EE_eq}), which causes the EE to decrease as the number of users increases. Apart from considering VNF migration, the point in the results is that the multi-agents methods, i.e., MAHHCDA and MADDPG, fall behind the single-agent ones, i.e., HHCDA and DDPG. There is explanation for these results, as in the single-agent methods the environment is fully observable to the agent, therefore, there is a centralized intelligence which can achieve the near optimal policy for VNF placement and recourse allocation among users. On the other hand, in multi-agents scenarios, since each agent has limit observation of the environment, i.e., it's position and some of the users' positions and requests, therefore they cannot achieve the near optimal policy. It should be noted that, multi-agent method can outperform the single-agent one by allowing each agent have full observation of the environment which would increase the method complexity. Another important result is considering hybrid actions in HHCDA compared to DDPG based method. Since there are some discrete actions as VNF placement and migration in the environment which is dependent to the UAVs and users positions, as well as the users' requests, our proposed HHCDA method selects the near optimal VNF plcament and migration actions in a hierarchical manner. In other words, the HHCDA agent decides to perform VNF placement and migration based on the UAVs movement, power and bandwidth allocation policy which theorically introduce the HHCDA agent more insight to select the near optimal VNF placement and migration. In other words, the HHCDA agent first selects the power allocation and the UAVs movements, then selects the VNF placement adn migration actions. 
	\begin{figure}
		\begin{center} 
			\subfigure[]{\label{mobility_1}\includegraphics[width=5cm]{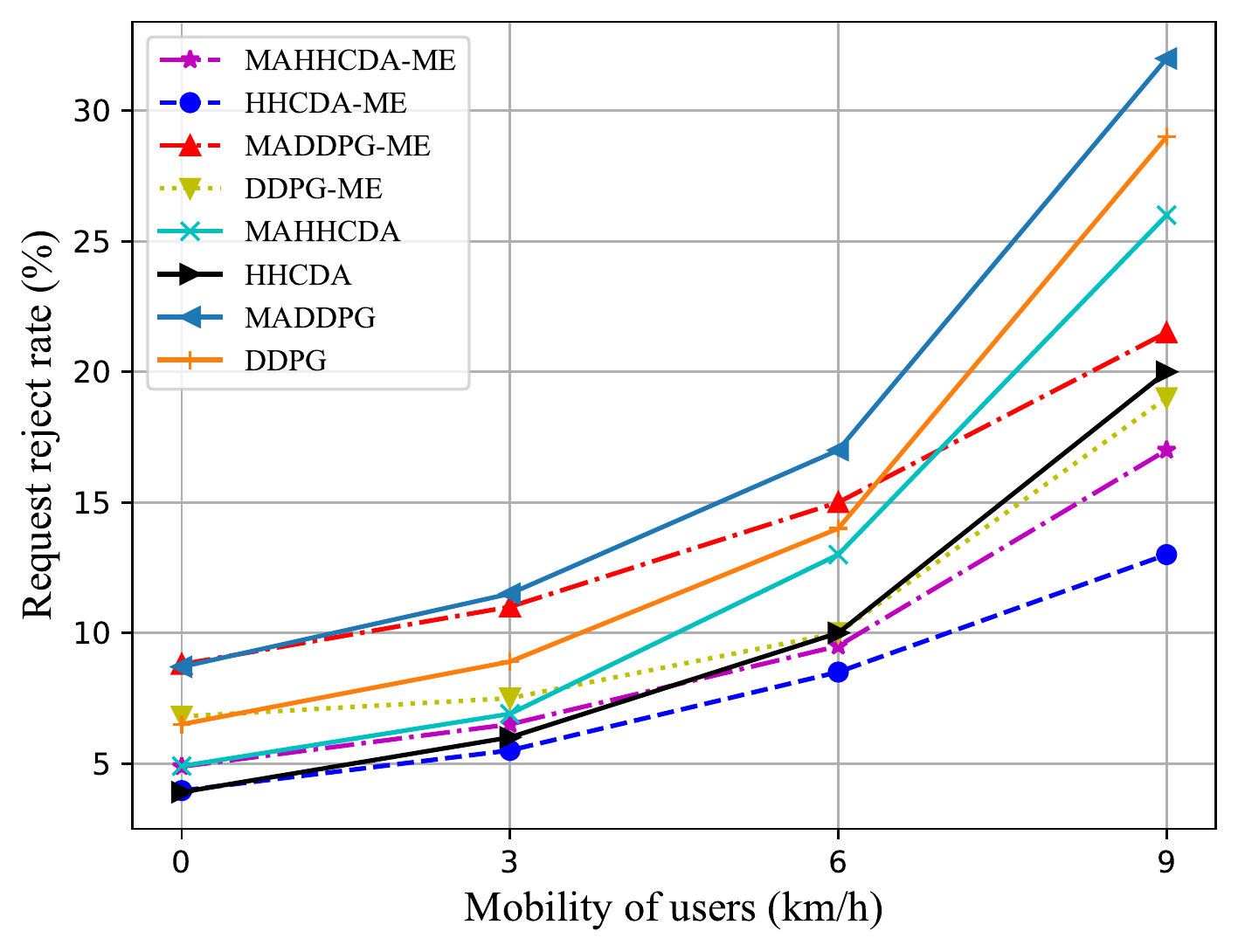}}
			\subfigure[]{\label{mobility_2}\includegraphics[width=5cm]{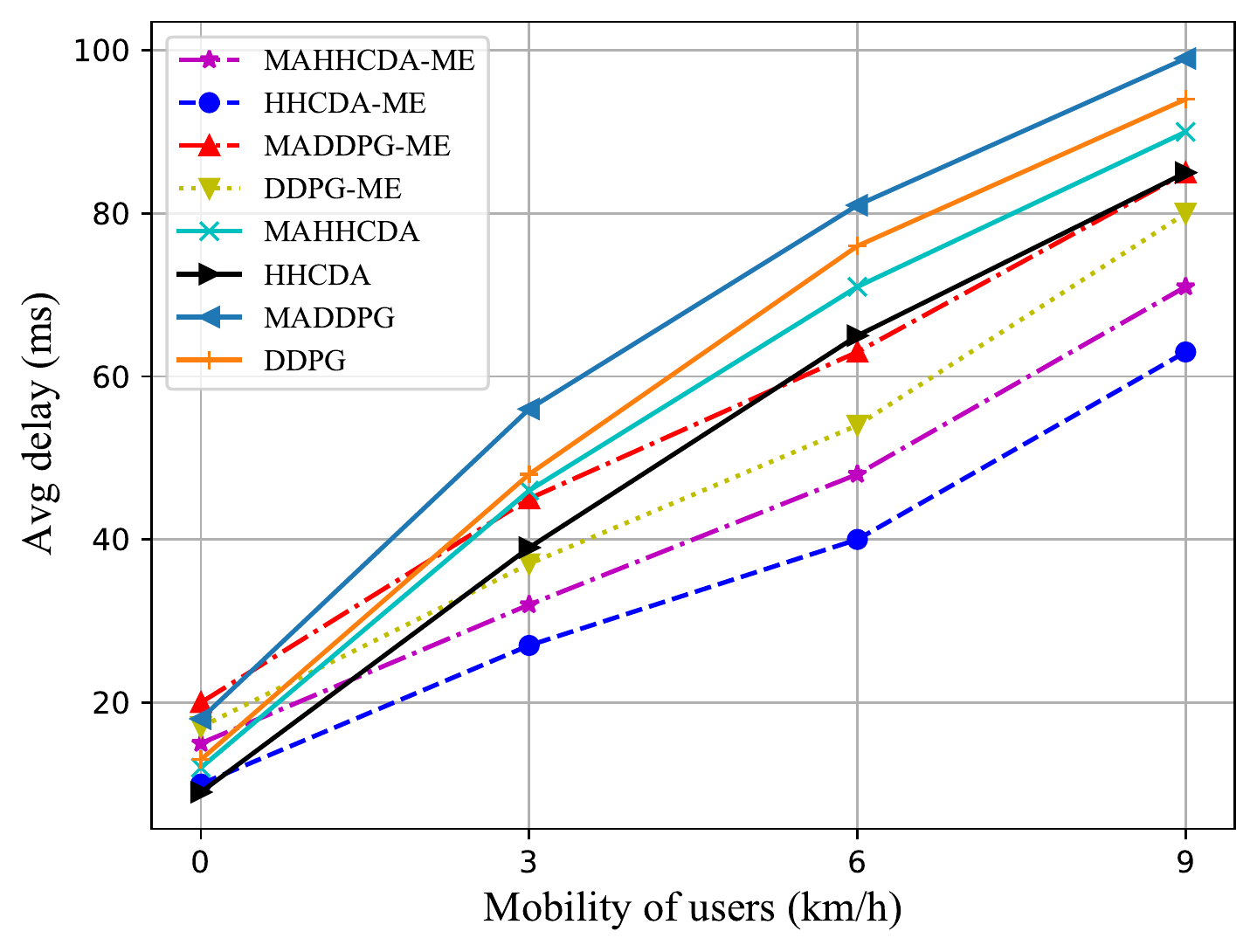}}
			\subfigure[]{\label{mobility_3}\includegraphics[width=5cm]{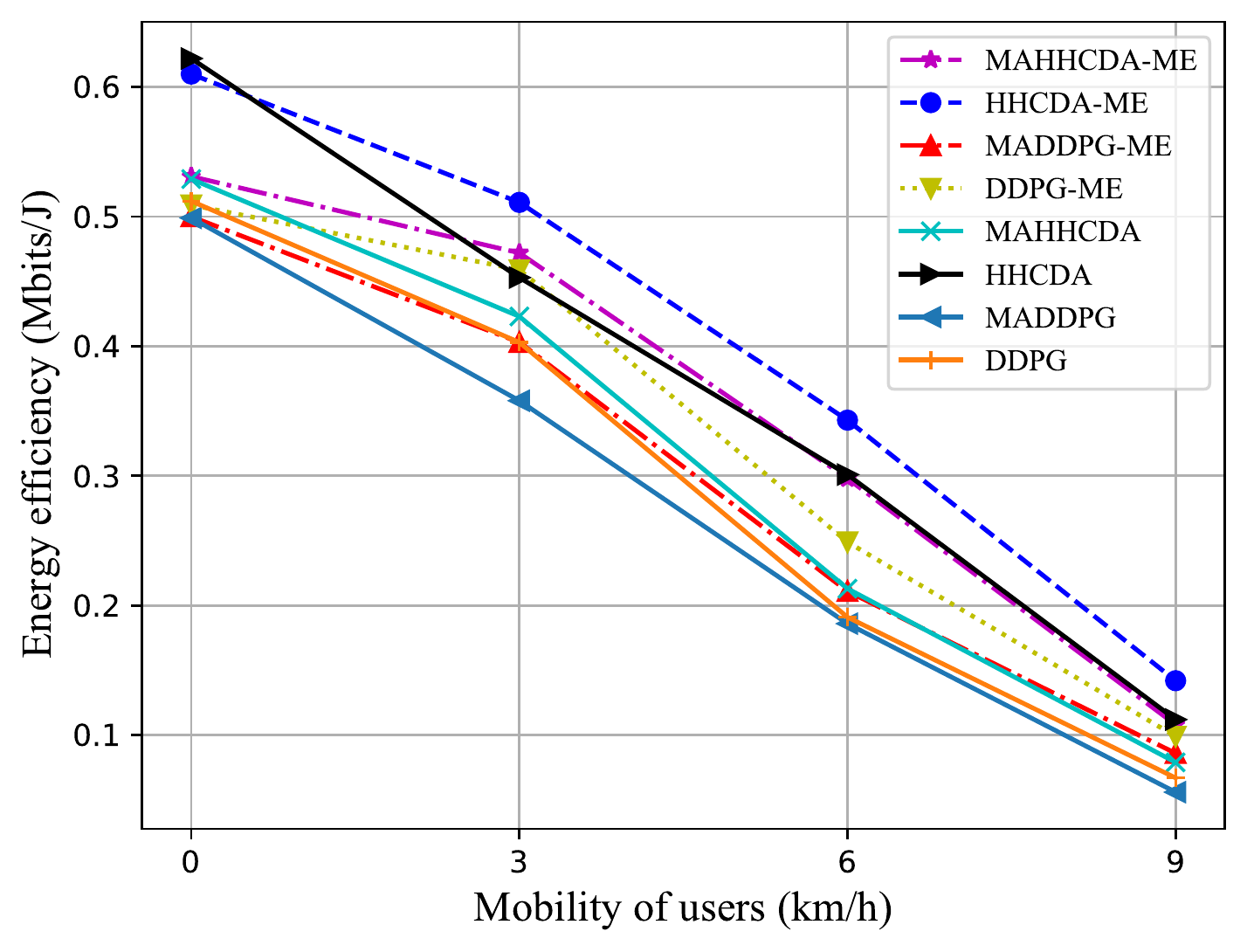}}
			\caption{Impacts of the mobility of users on (a) the RRR, (b) the average delay, and (c) the EE for different DRL methods with VNF migration and without VNF migration.} \label{mobility_impacts}
		\end{center}
	\end{figure}

	\begin{figure}
	\begin{center} 
		\subfigure[]{\label{mobility_1}\includegraphics[width=5cm]{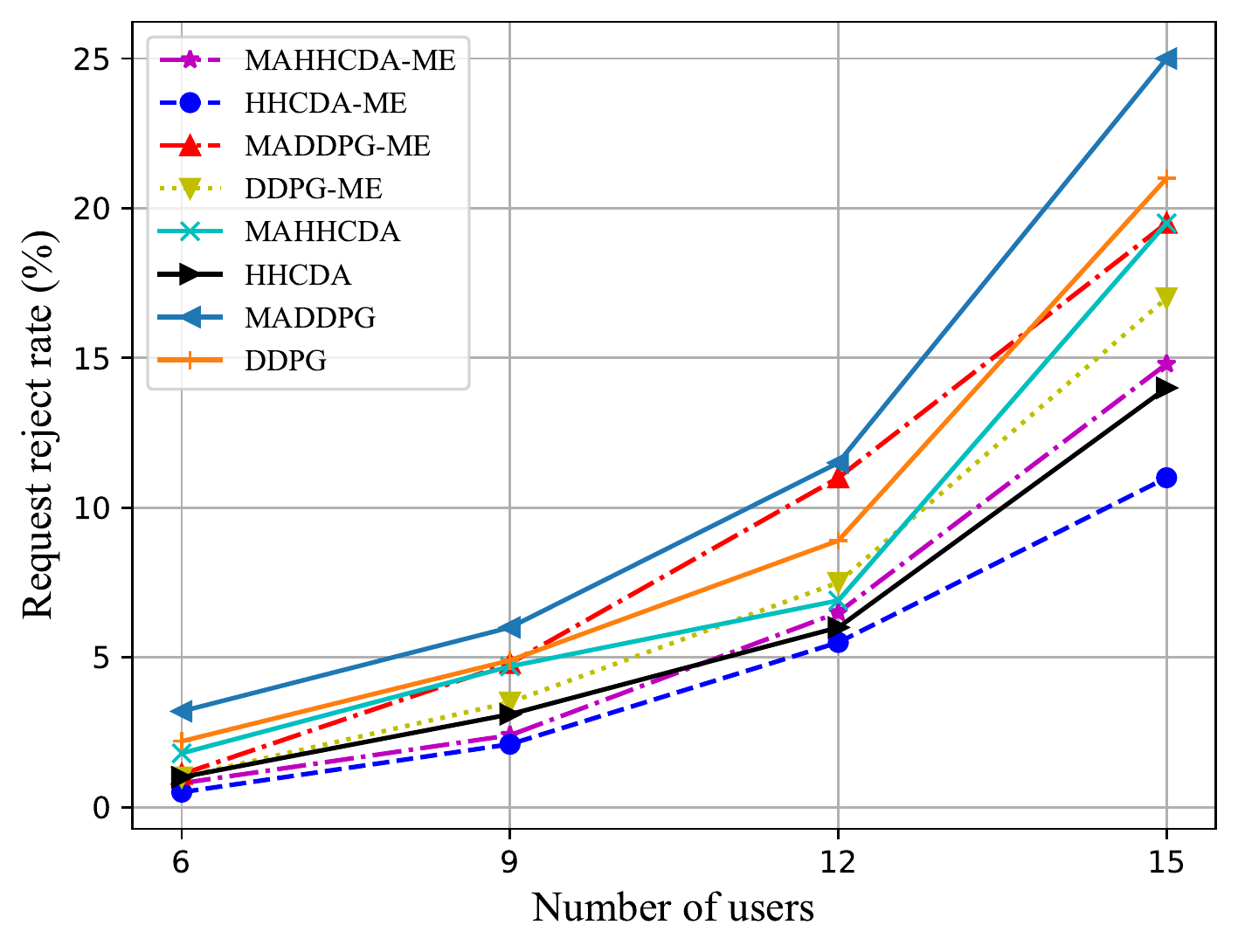}}
		\subfigure[]{\label{mobility_2}\includegraphics[width=5cm]{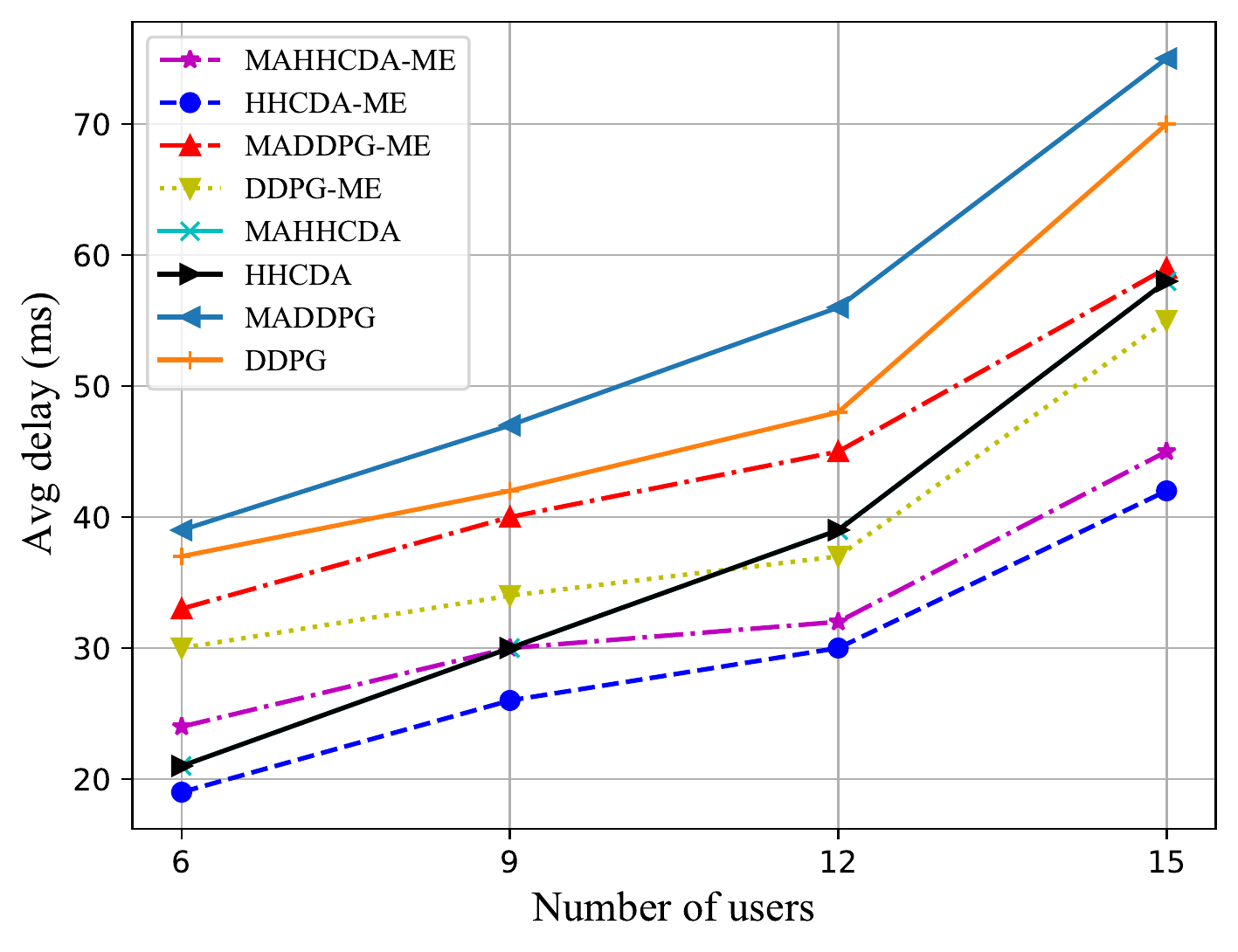}}
		\subfigure[]{\label{mobility_3}\includegraphics[width=5cm]{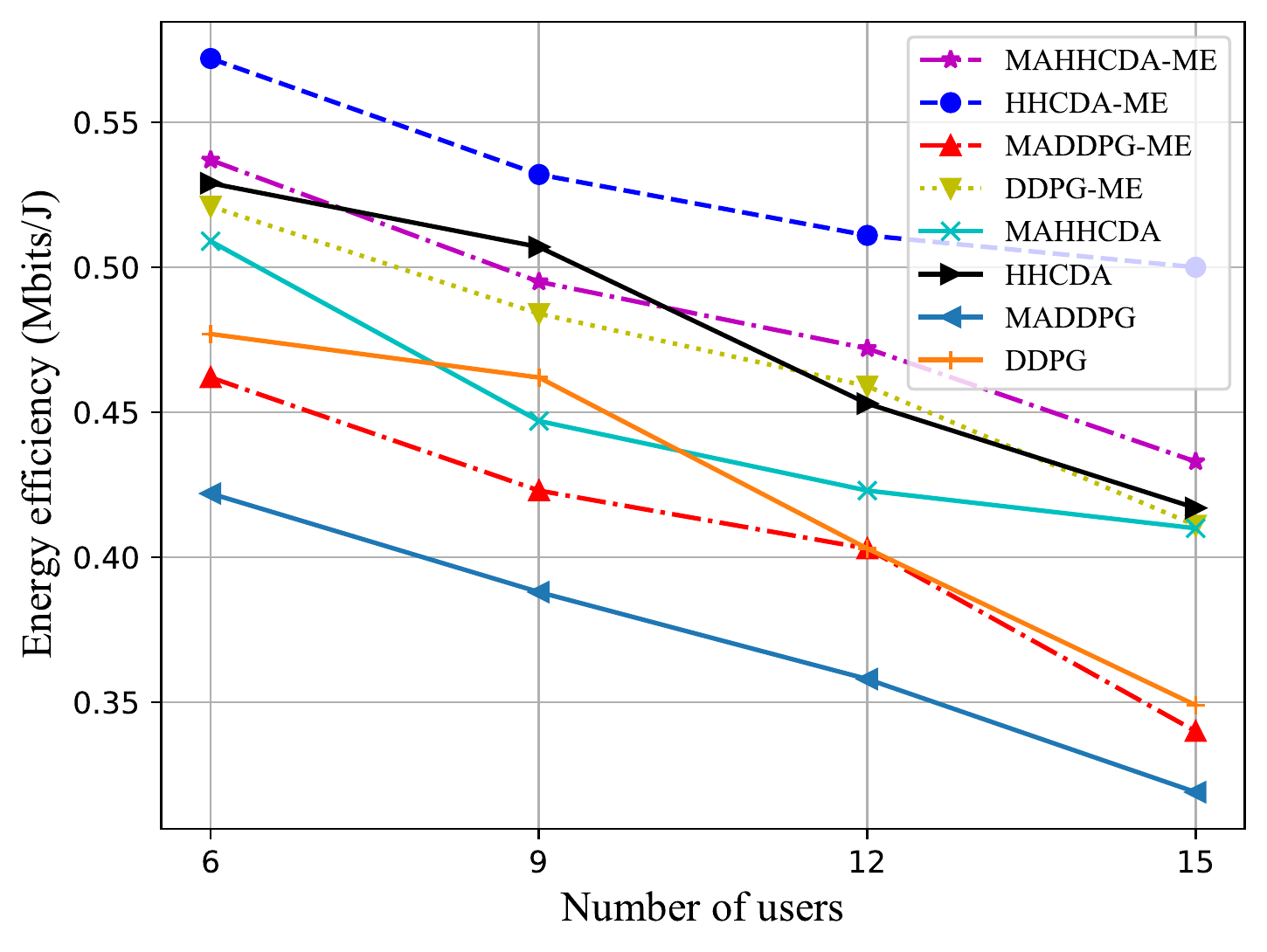}}
		\caption{Impacts of the number of users on (a) the RRR, (b) the average delay, and (c) the EE for different DRL methods with VNF migration and without VNF migration.} \label{users_impacts}
	\end{center}
\end{figure}

	\begin{table}[h]\label{table4}
		\small 
		\setlength{\tabcolsep}{38pt}
		\caption{Performance and Complexity Comparison versus DDPG method} 
		\centering          
		\begin{tabular}{|l  | l|l|} 
			\hline     
			Method & Relative Performance& Complexity\\
			\hline
			HHCDA & 65\% & 40\%\\
			MAHHCDA & 50\%& 30\%\\
			MADDPG& -25\% & -30\%\\
			\hline
		\end{tabular}%
	\end{table}
	The performance gain and complexity analysis of our DRL method versus DDPG and MADDPG are depicted in Table. \ref{table4}. The positive and negative values for the relative performance illustrate superiority and inferiority of the methods compared to DDPG, respectively, while for complexity, they indicate more complexity and less complexity compared to DDPG, respectively. Although our proposed HHCDA has more complexity,  its performance gain is higher (65\%) compared to DDPG method. On the other hand, multi-agent scheme of our proposed HHDCA method, has less complexity because the size of state space and action space are smaller compared to single-agent ones. In other words, in multi-agent method the state space and action space are divided into multiple smaller state and action spaces between agents, therefore the computational complexity is decreased.
	\section{Conclusion}
	In this paper, we investigated a dynamic resource allocation, trajectory planning, VNF placement, and scheduling framework for an UAV assisted network to support diverse services with different QoS requirements. We proposed a \textcolor{black}{hybrid action} DRL-based framework, \textcolor{black}{called} HHCDA, to
	facilitate efficient resource allocation and network management. Simulation results showed that our
	proposed migration enabled-scheme can effectively outperforms the state-of-the-art schemes and satisfies QoS requirements for high mobility users, especially when the number of users is large in the network. Moreover, for the implementation in large-scale UAV assisted networks, we developed a distributed, low-overhead, and low-complexity multi-agent scheme.
	
.
\ifCLASSOPTIONcaptionsoff
  \newpage
\fi



%
%


\bibliography{citation_AI-UAV-NFV}	
\bibliographystyle{ieeetran}
%




\end{document}